\documentclass[11pt,preprint]{aastex}
\usepackage{times}

\newcommand{\xmm}{{\it XMM--Newton}}
\newcommand{\suzaku}{{\it Suzaku}}

\newcommand{\Msun}{\hbox{$\rm\thinspace M_{\odot}$}}
\newcommand{\errUD}[2]{\ensuremath{^{+#1}_{-#2}}}
\newcommand{\logxi}{erg cm s$^{-1}$}
\newcommand{\nh}{cm$^{-2}$}
\newcommand{\nhsym}{N_{\mbox{\scriptsize H}}}
\newcommand{\ls}
{\mathrel{\hbox{\rlap{\hbox{\lower4pt\hbox{$\sim$}}}\hbox{$<$}}}}
\newcommand{\gs}
{\mathrel{\hbox{\rlap{\hbox{\lower4pt\hbox{$\sim$}}}\hbox{$>$}}}}
\def\Msun{\hbox{$\rm ~M_{\odot}$}}

\received{}
\begin{document}
\title{Discovery of Broad Soft X-ray Absorption Lines from the Quasar Wind in PDS 456}
\shorttitle{Absorption lines in PDS 456.}
\shortauthors{Reeves et al.}
\author{J. N. Reeves\altaffilmark{1,2}, V. Braito\altaffilmark{1,3},
 E. Nardini\altaffilmark{2}, E. Behar\altaffilmark{4,5}, P. T. O'Brien\altaffilmark{6}, 
F. Tombesi\altaffilmark{5,7}, T. J. Turner\altaffilmark{8}, M. T. Costa\altaffilmark{2}}

\altaffiltext{1}{Center for Space Science and Technology, 
University of Maryland Baltimore County, 1000 Hilltop Circle, Baltimore, MD 21250, USA; email jreeves@umbc.edu}
\altaffiltext{2}{Astrophysics Group, School of Physical and Geographical Sciences, Keele 
University, Keele, Staffordshire, ST5 5BG, UK; j.n.reeves@keele.ac.uk}
\altaffiltext{3}{INAF - Osservatorio Astronomico di Brera, Via Bianchi 46 I-23807 Merate (LC), Italy}
\altaffiltext{4}{Dept of Physics, Technion, Haifa 32000, Israel}
\altaffiltext{5}{Department of Astronomy, University of Maryland, College Park, MD 20742, USA}
\altaffiltext{6}{Dept of Physics and Astronomy, University of Leicester, University Road, Leicester LE1 7RH, UK}
\altaffiltext{7}{Astrophysics Science Division, NASA/Goddard Space Flight Center, Greenbelt, MD
20771}
\altaffiltext{8}{Department of Physics, 
University of Maryland Baltimore County, 1000 Hilltop Circle, Baltimore, MD 21250, USA}

\begin{abstract}

High resolution soft X-ray spectroscopy of the prototype accretion disk wind 
quasar, PDS 456, is presented. Here, the \xmm\ RGS spectra are analyzed from 
the large 2013--2014 \xmm\ campaign, consisting of 5 observations of approximately 
100\,ks in length. During the last observation (hereafter OBS.\,E), the quasar 
is at a minimum flux level and broad absorption line 
profiles are revealed in the soft X-ray band, with typical velocity 
widths of $\sigma_{\rm v}\sim 10,000$\,km\,s$^{-1}$. During a period of higher flux 
in the 3rd and 4th observations 
(OBS.\,C and D, respectively), a very broad absorption trough is also present above 
1\,keV.  From fitting the absorption lines with models of photoionized absorption 
spectra, the inferred outflow velocities lie in the range $\sim 0.1-0.2c$. 
The absorption lines likely originate from He and H-like neon and L-shell iron at these energies.
Comparison with earlier archival data of PDS 456 also reveals similar 
absorption structure near 1 keV in a 40\,ks observation in 2001, and generally 
the absorption lines appear most apparent when the spectrum is more absorbed 
overall. The presence of the soft X-ray broad absorption lines is also 
independently confirmed from an analysis of the \xmm\ EPIC spectra below 2\,keV. 
We suggest that the soft X-ray absorption profiles could be associated with a lower 
ionization and possibly clumpy phase of the accretion disk wind, where the latter 
is known to be present in this quasar from its well studied iron K absorption 
profile and where the wind velocity reaches a typical value of 0.3$c$.

\end{abstract}


\section{Introduction}

The masses of supermassive black holes (SMBHs, with $M_{\rm BH}=10^{6}-10^{9}M_{\odot}$) 
are known to correlate with galaxy bulge mass (Magorrian et al. 1998) and even more 
tightly with the stellar velocity dispersion on kpc scales; the so-called $M-\sigma$ relation 
(Ferrarese \& Merritt 2000, Gebhardt et al. 2000). This implies a co-evolution between SMBHs 
and their host galaxies, although the exact mechanism linking the two remained unclear. 
A potential mechanism responsible for this co-evolution came with the discovery of 
extremely energetic outflows from the black holes powering the most luminous Active 
Galactic Nuclei (AGN) and quasars (Pounds et al. 2003; Reeves et al. 2003; Chartas 
et al. 2002, 2003; Tombesi et al. 2010; Gofford et al. 2013). 
At high redshifts, such winds would have provided the necessary mechanical feedback 
that both controlled the formation of stellar bulges and simultaneously regulated SMBH growth, 
leaving the observed  $M-\sigma$ relation as a record of the process (Silk \& Rees 1998; 
King 2003; Di Matteo et al. 2005).

In the local Universe ($z < 0.3$), the most powerful and best characterized 
X-ray wind observed so far in an AGN is hosted by PDS\,456, a nearby ($z = 0.184$) 
radio-quiet quasar identified only in 1997 (Torres et al. 1997). PDS\,456 is a 
remarkable object in many respects. The optical and near-infrared spectra show 
Balmer and Paschen lines with broad wings (full-width at zero intensity of 
$>$30,000 km\,s$^{-1}$; Simpson et al. 1999), while in the $HST$/STIS UV spectrum 
the C\,\textsc{iv}\,$\lambda1549$\,\AA~emission line is blueshifted by $\sim$5,000 
km\,s$^{-1}$, and a tentative absorption trough extends from $\sim$14,000 to 
24,000 km\,s$^{-1}$ bluewards of the Ly$\alpha$ rest-frame energy (O'Brien et al. 
2005). The bolometric luminosity of PDS\,456, of the order of 
$L_{\rm bol}=10^{47}$\,erg\,s$^{-1}$ (Reeves et al. 2000; Yun et al. 2004), 
is more typical of a source at the peak of the quasar epoch ($z \sim 2$--3), 
when AGN feedback is thought to have played a major role in the evolution of 
galaxies. No direct measurement is available for the mass of the central 
black hole, but this can be estimated from the SMBH/host galaxy scaling relations 
to be $\sim 1$--$2 \times 10^9 \Msun$ (Nardini et al. 2015), thus implying 
that the black hole in PDS\,456 is accreting at a substantial fraction of the 
Eddington rate. Under these physical conditions, the photon momentum flux can 
contribute to the driving of massive accretion-disk winds (several ${\rm M}_{\odot}$ 
yr$^{-1}$), provided that the nuclear environment is sufficiently opaque to the 
continuum radiation (e.g. King 2010; see also Hagino et al. 2015).

The clear presence of strong absorption above 7 keV in PDS\,456 was revealed 
by a short (40-ks) \xmm\ observation in 2001. If attributed to iron K-shell 
absorption, such a feature would have arisen in a high velocity outflow, requiring 
a large column density of highly ionized matter (Reeves et al. 2003). An unusual 
absorption trough was also found in the soft X-ray band near 1 keV (see also 
Behar et al. 2010). The detection of fast Fe K absorption has been confirmed in 
subsequent {\it Suzaku} campaigns in 2007, 2011 and 2013, which implied an outflow 
velocity of $\sim0.3\,c$ (Reeves et al. 2009; Reeves et al. 2014; Gofford et al. 2014). 
The broad, blueshifted emission and absorption profiles in the UV could be then 
potentially associated with the decelerating phase of the wind out to large scales.

PDS\,456 was observed again with \xmm\ in a series of five observations between 
August 2013 and February 2014, with the first four sequences carried out over about 
a month and the last one six months later, in order to sample the spectral variations 
over different time scales. All the observations were complemented by the simultaneous 
high-energy spectra collected by {\it NuSTAR}, which provide a valuable broadband 
view that extends from the optical/UV to the hard X-rays. In this campaign, Nardini 
et al. (2015) were able to detect a persistent P-Cygni like profile from highly 
ionized iron, thus establishing the wide-angle character of the disk wind in PDS\,456.
This proved to be important in accurately determining the overall energy budget of 
the outflow in PDS\,456, made possible through establishing that the overall wind 
solid angle was a significant fraction of $4\pi$~steradian. Thus the inner disk wind 
expels matter at rates close to Eddington, with a kinetic power a significant fraction 
of the bolometric output of the quasar, comfortably exceeding the typical values 
thought to be significant for quasar mode feedback (Hopkins \& Elvis 2010). In many 
respects, PDS\,456 strongly resembles the two fast outflows recently measured in two other 
luminous obscured quasars, IRAS\,F11119+3257 (Tombesi et al. 2015) and Mrk\,231 
(Feruglio et al. 2015). Here the initial accretion disk wind is likely critical 
for evacuating matter from the central regions of these systems during their post 
merger phases, with the eventual fate of the gas seen at large ($\sim$\,kpc) scales 
through massive ($\sim1000$\,\Msun\,yr$^{-1}$) energy conserving molecular outflows.

However to date there have been very few detections of fast AGN outflows in the soft X-ray band (see 
Pounds 2014; Longinotti et al. 2015; Gupta et al. 2013, 2015 for some possible recent 
examples). Here we present the soft X-ray spectroscopy of PDS\,456 obtained from the 
\xmm\ Reflection Grating Spectrometer (RGS; den Herder et al. 2001) as part of the 
extended 2013--2014 campaign. In the subsequent analysis we will show the 
detection of broad soft X-ray absorption profiles in PDS\,456, which may be 
potentially associated to the fast wind measured in the iron K band. 
In Section\,2, we outline the \xmm\ observations 
and data reduction, in Section\,3 the overall form of the mean RGS spectrum is discussed, 
while Section 4 presents the detection of the soft X-ray absorption profiles in the 
individual RGS observations. In Section 5 photoionization modeling of the putative wind 
in the soft X-ray band is presented, while Section 6 discusses the origins of the 
absorbing gas from a clumpy, multi phase accretion disk wind. Throughout this work 
we assume the concordance cosmological values of H$_{\rm 0}$=70\,km\,s$^{-1}$\,Mpc$^{-1}$ 
and $\Omega_{\Lambda_{\rm 0}}=0.73$, and errors are quoted at 90\% confidence 
($\Delta\chi^{2}=2.7$) for one parameter of interest. In the spectral analysis, we 
adopt a conversion between energy and wavelength of $E = (12.3984$ \AA/$\lambda)$\,keV.

\section{XMM-Newton Observations of PDS 456}

Five \xmm\ observations of PDS 456, of at least 100\,ks in duration, were performed 
over a six month time period in 2013--2014. The EPIC-pn and MOS instruments were 
operating in Large Window mode and with the Thin filter applied. The observations are 
listed in Table\,1 and are described in more detail in Nardini et al. (2015). The first 
four observations (hereafter OBS.\,A--D) were performed within a four week period (with 
separations of $4-10$\,days between observations), while the fifth observation (OBS.\,E) 
was performed about 5 months later and caught the AGN in the lowest flux state of all the 
campaign (see Table\,1). The \xmm\ data have been processed and cleaned using the Science 
Analysis Software (\textsc{sas} v14.0.0), and analyzed using standard software packages 
(\textsc{ftools} v6.17, \textsc{xspec} v12.9). 

For the scientific analysis of this paper we concentrated on the RGS data, which have 
the highest spectral resolution in the soft X-ray band, and we compared the RGS results
with the EPIC MOS and pn data for consistency. The RGS data have been reduced using the 
standard \textsc{sas} task {\it rgsproc} and the most recent calibration files. High 
background time intervals have been filtered out applying a threshold of 0.2 counts s$^{-1}$ 
on the background event files. OBS.\,A was affected by severe telemetry issues and about 
28 ks into the observation, even if the data were still recorded, the RGS telemetry 
became corrupted. The main effect is not a simple loss of the data but an incorrect 
count rate for each of the RGS spectra. As it was not possible to recover the ODF files 
and as the subsequent RGS spectra were not suitable for analysis, hereafter we concentrate 
only on OBS.\,B--E from this campaign. Nonetheless during OBS.\,A, from an analysis of the 
unaffected EPIC spectra, PDS\,456 was caught in a bright and less obscured state, which 
was mainly featureless in the soft X-ray band (see Nardini et al. 2015).

For each of the remaining four observations we first checked that the RGS\,1 and 
RGS\,2 spectra were in good agreement, typically to within the 3\% level, 
and we subsequently combined them with the 
\textsc{sas} task {\it rgscombine}. The spectra were analyzed over the 0.45--2.0\,keV 
energy range; below 0.45\,keV the spectra are noisy due to the Galactic absorption column 
towards PDS\,456. During these observations the two RGS collected typically $\sim 
8600 - 13600$  net counts per observation (see Table~1 for details). Two of the 
observations (OBS.\,C and OBS.\,D), which were separated by only 4 days and show 
a very similar spectral shape and count rate, were further combined using 
{\it rgscombine} into a single spectrum (OBS.\,CD). Finally, in order to investigate 
the mean spectral properties with a higher signal to noise spectrum, we combined all 
the four available observations obtained during this campaign.

In the initial spectral fitting, we first adopted a spectral binning of 
$\Delta\lambda=0.1$\,\AA\ for the RGS spectra, which just slightly under-samples 
the FWHM spectral resolution of $\Delta\lambda=0.06-0.08$\,\AA. At this binning, 
the spectra have $>20$ counts per bin and thus $\chi^{2}$ minimization was used 
for all the spectral fitting. Note that in the subsequent analysis we also 
considered a finer binning  (of $\Delta\lambda=0.05$\,\AA) in order to search 
for any narrow features in the combined RGS\,1+2 spectra, as well as a coarser 
binning (of $\Delta\lambda=0.2$\,\AA) for modeling the broad spectral features 
present in some of the individual sequences.

The \xmm\ EPIC data were filtered for high background time intervals, which yields net 
exposure times as listed in Table 1. For the analysis we concentrated on the MOS spectra, 
as they offer a better spectral resolution in the soft band compared to the pn data; 
however, we did check the pn spectra for consistency in each observation. The MOS\,1+2 
source and background spectra were extracted using a circular region with a radius of 
$35''$ and two circular regions with the same radius, respectively. Response matrices 
and ancillary response files at the source position were created using the \textsc{sas} 
tasks \textit{arfgen} and \textit{rmfgen}. We then combined for each exposure the MOS\,1 
and MOS\,2 spectra, after verifying that the individual spectra were consistent. As for 
the RGS data, after checking that the MOS spectra for OBS.\,C and OBS.\,D were similar, 
we combined them into a single spectrum.

The MOS source spectra were then binned with constant energy intervals of 15 eV. This 
oversamples the spectral resolution of the EPIC MOS detectors below 2 keV, where the FWHM 
is $\sim 50$\,eV (60 eV) at 0.5 keV (1 keV), by only  factor of 3--4. We note that 
adopting this binning allows us to simultaneously fit both the MOS and the RGS data with 
a similar sampling. The source is bright enough to collect between $0.54-0.79$\,counts\,s$^{-1}$ 
for the combined MOS\,1+2 spectra over the $0.5-2$ keV band.

In order to study the long term spectral variability of the PDS\,456 soft X-ray 
spectra, the RGS and EPIC spectra from the 2001 and 2007 archival observations were 
also extracted, (see Table\,1). The data were reduced in an identical way as to the 
2013--2014 observations. The 2001 observation consisted of a single exposure of 
$\sim 40$\,ks in duration, but at a higher overall count rate compared to the 
2013--2014 observations, while the 2007 observations consisted of two sequences 
over two consecutive \xmm\ orbits. These latter two 2007 spectra were consistent 
in spectral shape, with just a simple offset in flux between them, so they were 
combined to give a single RGS spectrum from 2007.

\section{Initial Mean RGS Spectral Analysis}

Initially we combined all of the 2013--2014 RGS observations of PDS\,456 (i.e., OBS.\,B 
through to OBS.\,E), in order to construct a single, time-averaged spectrum. Although 
the spectrum of PDS\,456 is time variable, the idea is to have an initial parameterization 
of both the continuum and local Galactic absorption before proceeding with the analysis 
of the individual sequences. The mean RGS spectrum was fitted with a simple power-law 
plus blackbody continuum form, where the blackbody is required to account for the soft 
excess towards lower energies seen in this AGN. The mean spectrum was initially binned 
in constant wavelength bins of $\Delta\lambda=0.1$\,\AA\ over the 0.45--2.0\,keV (or 
6.2--28.0\,\AA) range. 
A Galactic absorption column from 21-cm measurements of 
$N_{\rm H}=2.4\times10^{21}$\,cm$^{-2}$ (Dickey \& Lockman 1990; Kalberla et al. 2005)
was adopted at first, modeled with the \texttt{tbnew} ISM absorption model in 
{\sc xspec} using the cross--sections and abundances of Wilms et al. (2000). 
However, in subsequent fits, the Galactic $N_{\rm H}$ value was allowed to vary, 
along with the relative abundance of Oxygen (compared to H) in order to ensure a 
good fit in the region around the neutral O edge.

This simple continuum form returned a photon index of $\Gamma=2.1\pm0.2$, with a blackbody 
temperature of $kT = 102^{+15}_{-10}$\,eV, while the Galactic column was found to be in 
excess of the 21-cm value with $N_{\rm H}=(3.9\pm0.6)\times10^{21}$\,cm$^{-2}$. 
The resulting fit is shown in Figure\,1, where the Galactic absorption model reproduces 
well the structure observed around the neutral O edge at $\sim$0.5\,keV. The 
relative O abundance was found to be higher by a factor of $A_{\rm O}=1.27\pm0.12$ 
compared to the O/H abundance of $4.9\times10^{-4}$ reported in Wilms et al. (2000). 
The absorbed 0.5--2.0\,keV soft X-ray flux resulting from this model is 
$2.14\times10^{-12}$\,erg\,cm$^{-2}$\,s$^{-1}$. Note that if instead the Solar 
abundances of Grevesse \& Sauval (1998) 
are adopted, which present a higher abundance of O, then the relative O abundance is 
consistent with Solar ($A_{\rm O}=0.95\pm0.09$). While this model reproduces the overall 
shape of the soft X-ray continuum, the fit statistic is relatively poor with a resulting 
reduced chi-squared of $\chi^{2}/\nu = 287.7/205$, suggesting the presence of additional 
spectral complexity. An almost identical fit is obtained 
if a somewhat different form for the soft excess is used, e.g. a power-law plus 
Comptonized-disk spectrum. 

The mean spectrum was also binned at a finer resolution of $\Delta\lambda=0.05$\,\AA\ 
per bin, in order to better sample the resolution of the RGS over the spectral range of 
interest. The motivation for this is to check whether there is any signature of a warm 
absorber from low velocity, low to moderately ionized outflowing gas, which is commonly 
detected in the high resolution soft X-ray spectra of nearby Seyfert 1 galaxies 
(Kaastra et al. 2000, Sako et al. 2001, Kaspi et al. 2002). 
Furthermore 
any such low velocity (and distant) component of a warm absorber would not be expected 
to strongly vary between the observations.
Thus with the high signal to 
noise of the mean 464\,ks net exposure, any narrow absorption (or emission) lines from 
distant and less variable 
photoionized gas should be readily apparent in the spectrum close to the expected 
rest-frame energies of the strongest atomic lines in the soft X-ray band. 
The residuals 
to the above absorbed power-law plus blackbody fit are shown in Figure 2, which is plotted 
in the rest frame of PDS\,456. It is clear that upon inspection of the spectra, no strong 
narrow absorption or emission lines are observed close to the expected positions of the 
strongest lines, such as the He-$\alpha$ and Ly-$\alpha$ resonance lines of O, Ne, Mg or Si, 
or at the energy of the Fe M-shell UTA. 

In order to place a formal limit on the column of any warm absorber towards PDS\,456, 
the spectrum was modeled with an \textsc{xstar} absorption grid with a turbulence 
velocity of $\sigma=300$\,km\,s$^{-1}$, adopting Solar abundances (Grevesse \& Sauval 1998) 
and allowing for a modest outflow velocity of up to 5,000\,km\,s$^{-1}$. The above baseline 
continuum was used. 
The addition of this warm absorber component resulted in little improvement in the 
overall fit. Formally we can place an upper-limit on the column density of between 
$N_{\rm H} < 0.6\times10^{20}$\,cm$^{-2}$ and 
$N_{\rm H} < 1.7\times10^{20}$\,cm$^{-2}$ for an ionization parameter\footnote{The 
ionization parameter is defined as $\xi=L_{\rm ion}/n_{\rm e} r^{2}$, where $n_{\rm e}$ 
is the electron density, $r$ is the radial distance between the X-ray emitter and the 
ionized gas and $L_{\rm ion}$ is the ionizing luminosity over the 1--1000\,Rydberg range. 
The units of $\xi$ are subsequently erg\,cm\,s$^{-1}$.} in the range 
from $\log \xi=0-2$, which covers the typical range of ionization seen in the soft 
X-ray warm absorbers in Seyfert 1s. Thus the presence of a low velocity, low 
ionization warm absorber appears ruled out in PDS\,456.

Although no conventional signature of a warm absorber, in the form of narrow absorption lines, 
appears in these RGS spectra, 
the finely binned mean spectrum does reveal broad residual structure over the energy 
range from 0.8--1.6\,keV. Indeed, fitting with the above baseline continuum results 
in a very poor fit, with $\chi^{2}/\nu = 573/406$ and a corresponding null hypothesis 
probability of $8.6\times10^{-8}$. 
In the following 
sections we investigate whether these broad features may represent the soft X-ray signature 
of the known fast wind towards PDS\,456 detected at iron K (e.g. Reeves et al. 2009; 
Nardini et al. 2015). 
In the subsequent spectral fitting we adopt the coarse spectral binning of 
$\Delta\lambda=0.2$\AA\ to model these broad profiles, as no narrow features are 
required by the data.

\section{Analysis of Individual RGS Sequences}
We then proceeded to investigate each of the RGS spectra collected during 
the 2013--2014 campaign. A first inspection of the individual RGS spectra 
shows that there are remarkable differences between OBS.\,B (upper and black 
data points in the upper panel of Figure~\ref{fluxed}) and OBS.\,E (blue data 
points). The 0.5--2 keV spectral shapes appear broadly similar, with some 
variability of the intensity of the emerging continuum. In addition, the 
appearance of a deep trough  at $\sim 1$ keV in OBS.\,E (observed frame) 
suggests that the variations could at least in part be caused by a variable 
soft X-ray absorber. On the other hand, the spectra obtained during OBS.\,C 
and OBS.\,D are almost identical (see Figure~\ref{fluxed} lower panel) and are 
intermediate in flux between the OBS.\,B and OBS.\,E. Thus in order to investigate 
the main driver of the soft X-ray spectral variability, we proceeded to fit each 
of the main spectral states (i.e. OBS.\,B, OBS.\,CD and OBS.\,E) independently. 
For the continuum we adopted the blackbody plus power-law continuum model 
found with the analysis of the mean RGS spectrum, allowing the $\nhsym$ of 
the Galactic absorption to vary, as well as the continuum parameters.

We found that, while for OBS.\,CD and OBS.\,E the photon index of the primary 
power law component is consistent with $\Gamma=2.1$, during OBS.\,B the spectrum 
requires a steeper power-law component ($\Gamma \sim 2.5\pm 0.2$). Statistically 
the fit is poor for most of the sequences when compared to the above continuum 
model, i.e. $\chi^2/\nu= 149.9/100 $ and $\chi^2/\nu= 170.2/100 $, for OBS.\,CD 
and OBS.\,E, respectively (see Table 2). In contrast, the continuum model can 
reproduce the overall shape of the RGS spectrum OBS.\,B ($\chi^2/\nu= 131.2/100$); 
indeed, as seen in  Figure~\ref{sequences} (top panel) OBS.\,B appears almost 
featureless in the RGS. 

However, the continuum model leaves clear residuals in the 0.9--1.3 keV energy 
range for OBS.\,CD and OBS.\,E. Broad and deep residuals are present in both of 
these spectra (Figure~\ref{sequences}, second and third panels), where three main 
absorption troughs are detected. During OBS.\,CD a very broad absorption feature 
is present near 1.2\,keV, while during OBS.\,E two absorption troughs emerge in 
the residuals at $\sim 1.0$\,keV and $\sim 1.17$\,keV, respectively. 
In particular, a closer inspection of the residuals of the OBS.\,E spectrum 
to the best fit continuum model (see Figure~\ref{pcyg}) unveils complex absorption 
structure around 1\,keV.

\subsection{Modeling and Identification of the Individual Profiles}

To parameterize these profiles we added several Gaussian absorption and emission 
lines to the baseline continuum model. We imposed an initial selection criteria, 
whereby any line in the RGS spectra was added to the baseline model only
if its addition yielded an improvement in the fit statistic 
of at least $\Delta \chi^2= 9.2$ (for 2 interesting parameters) or $\Delta \chi^2= 11.3$
(for 3 interesting parameters), equivalent to the 99\% significance level in $\chi^{2}$ 
statistics. Furthermore we impose an additional constraint, whereby any individual 
line in the RGS should be independently 
confirmed corresponding to at least the same improvement in $\Delta\chi^{2}$ in the MOS (see Section\,4.3). 
Thus requiring a possible line detection in the RGS to be independently confirmed 
in the MOS, at a self consistent energy and flux, 
reduces the likelihood that the line in question is spurious due to photon noise.
Subsequently the parameters of these Gaussian lines selected in the RGS are listed in Table~\ref{Gauss}, 
as well as the F-test probabilities associated with the addition of each line to the baseline model. 
Their statistical significances, which are typically confirmed at the $5\sigma$ level or higher once 
the MOS data are accounted for, are evaluated more throughly in the Appendix. 

As noted above, in OBS.\,CD we detected a strong ($EW=42 \pm 13$ eV) and very broad 
($\sigma \sim 100$ eV) absorption line at $E=1174\pm 38$ eV, with an apparent velocity 
width of $\sigma_{\rm v}=28000\errUD{13000}{9000}$\,km s$^{-1}$. We attempt to place 
a tentative identification on this absorption feature. The $\sim 1.2$\,keV absorption 
line could be identified with blue-shifted ($v_{\rm out} \sim 0.2\,c$) Ne\,\textsc{x} 
Ly$\alpha$ ($E_\mathrm{Lab}=1.022$\,keV), perhaps blended with iron L absorption in the 
ionization range from Fe\,{\textsc{xx--xxiv}}. Alternatively, associating this broad 
absorption line with a single feature, but with a smaller blueshift, would require an 
identification with the Fe\,\textsc{xxiv} $2s\rightarrow 3p$ at $E_\mathrm{Lab}=1.167$\,keV. 
However, this would imply a single very broad line seen in isolation, without a strong 
contribution from other Fe L lines, e.g. Fe\,\textsc{xxiii} at $E_\mathrm{Lab}=1.125$\,keV 
and lower ionization ions. 
Furthermore, as we will discuss later in 
Section 5, modeling this absorption profile with a grid of photoionized absorption 
spectra prefers the solution with a fast and highly ionized wind, noting that part 
of the large velocity width of the broad trough could be explained with a blend of 
these lines. Finally a narrower absorption line is also apparent at 846\,eV in 
OBS.\,CD, and although it is much weaker than the broad 1.2\,keV profile, it is 
also independently confirmed in the MOS (see Section 4.3). While the identification 
of this isolated line is uncertain, we note that if it is tentatively associated to 
absorption from O\,\textsc{viii} Ly$\alpha$ (at $E_\mathrm{Lab}=0.654$\,keV), its 
outflow velocity would be consistent with that found at iron K, at $\sim 0.3\,c$.

The appearance of complex soft X-ray absorption structure near 1\,keV in OBS\,E is perhaps 
not unexpected, especially as this RGS sequence appears the most absorbed and at the 
lowest flux of all the 2013--2014 \xmm\ data sets.
As shown in Figure\,\ref{pcyg}, two deep absorption lines (at $E\sim 1.01$\,keV and 
$E\sim 1.17 $\,keV) and an emission line at $E\sim 0.9$\,keV are formally required by 
the data (see Table\,2). 
Indeed, the addition of the two Gaussian absorption lines results in the fit 
statistic improving by $\Delta \chi^2=44.3$ for $\Delta\nu=5$. Assuming that the 
absorption line profiles have the same width, then the two lines are found to be 
resolved with a common width of $\sigma=41^{+14}_{-12}$\,eV (or 
$\sigma_{\rm v}=12000^{+4000}_{-3500}$\,km\,s$^{-1}$). 

One likely identification of the emission/absorption line pair in OBS.\,E at 
913\,eV and 1016\,eV may be with Ne\,\textsc{ix} (at $E_\mathrm{Lab}=0.905$ keV), 
which would then require the absorption line to be blueshifted with respect to the 
emission line component. Alternatively, if the absorption at 1016\,eV is separately 
associated with Ne\,\textsc{x} Ly$\alpha$ (at $E_\mathrm{Lab}=1.022$\,keV), then 
this would require little or no velocity shift. However, the second absorption 
line at 1166\,eV only requires a blueshift if it is associated with Ne\,\textsc{x} 
Lyman-$\alpha$, alternatively it could be associated to Fe\,\textsc{xxiv} $2s\rightarrow3p$ 
without requiring a blueshift.
Thus the precise identification of these 1\,keV absorption 
lines are difficult to determine on an ad-hoc basis from fitting simple Gaussian profiles.
Their most likely origin will be discussed further when we present the 
self consistent photoionization 
modeling of the OBS.\,E spectrum in Section\,5.3.
Nonetheless, regardless of their possible identification, the detection of the 
broad absorption profiles in the soft X-ray band in both OBS.\,E and OBS.\,CD may 
suggest the presence of a new absorption zone with somewhat lower velocity and/or 
ionization compared to the well-known ionized wind established at iron K.  \\

\subsection{Comparison to the 2001 and 2007 \textit{XMM--Newton} Observations}

Following the results of the 2013--2014 observations, we re-analyzed the RGS spectra 
collected during the past \xmm\ observations of PDS\,456, when the quasar was observed
in the two extreme states: a highly obscured one (2001; Reeves et al. 2003) and an 
unobscured state (2007; Behar et al. 2010). For the continuum model we again adopted 
the best fit found with the analysis of the mean RGS spectrum, allowing the $\nhsym$ 
of the Galactic absorption to vary as well as allowing the continuum parameters to 
adjust. For the 2001 observation we included an additional neutral partial covering 
absorber in the continuum model; indeed, during this observation, even though the 
intrinsic flux of PDS\,456 is relatively high (see Table\,1), the AGN appeared heavily 
obscured, with the presence of strong spectral curvature over the 1--10\,keV band 
(Reeves et al. 2003). This curvature is especially evident in the EPIC MOS data 
(see Section 4.3). Without this absorber the derived photon index, although poorly 
constrained, is extremely hard ($\Gamma\sim 1.5$), and its extrapolation above 2 keV     
lies well above the EPIC spectra. On the other hand, the 2007 spectrum required a 
steeper photon index ($\Gamma=2.4-2.5$), indicating a lack of intrinsic absorption. 

The residuals for the 2001 and 2007 RGS data to the baseline continuum model are 
shown in the two lower panels of Figure~\ref{sequences}. Similarly to the OBS.\,E 
spectrum during the 2013--2014 campaign, the 2001 observation tracks the presence 
of the highly ionized wind. 
The presence of absorption near 1\,keV in the 2001 observation was first noted 
in this data set by Reeves et al. (2003). A highly significant broad absorption 
profile is apparent in the residuals at $E=1061\pm 11$ eV, with an equivalent width 
of $EW\sim 37 $\,eV, while the improvement in the fit statistic upon adding this 
absorption line is $\Delta \chi^2/\Delta\nu=42.0/3$  (see Table 2). 
The profile is resolved with a width of $\sigma=42^{+13}_{-11}$\,eV (or 
$\sigma_{\rm v}=11900\errUD{3700}{3100}$\,km s$^{-1}$), as per the similar 
broad profiles in OBS.\,E. As discussed above, one plausible identification is 
with mildly blueshifted Ne\,\textsc{x} Ly$\alpha$ (again notwithstanding any 
possible contribution from iron L absorption). At face value this suggests the 
presence of a lower velocity zone of the wind, which is investigated further in 
Section\,5.
In contrast to the 2001 spectrum, the 2007 observations show little evidence for 
intrinsic absorption and appear to be featureless (see also Behar et al. 2010).  

\subsection{Consistency check with EPIC-MOS soft X-ray spectra}

In this section we perform a consistency check on the lines detected in the RGS 
with the EPIC MOS spectra. Indeed, given the strength and breadth of the absorption 
features detected in the RGS spectra, we may expect them to be detectable at the 
MOS CCD spectral resolution.  
We thus considered the combined MOS\,1+2 spectra for each observation and, while 
we fitted only the 0.5--2\,keV energy range where the soft X-ray features occur, 
we also checked that the best fit models provide a good representation of the overall 
X-ray continuum to higher energies. The continuum model found from the analysis of 
the RGS spectra was adopted, but again allowing the parameters to adjust. As noted 
above, the 2001 MOS observation also shows pronounced curvature between 1.2--2.0\,keV, 
which is apparent in the residuals of this spectrum in Figure~\ref{MOS}, and was 
accounted for by including a neutral partial covering absorber.

The residuals of the five MOS spectra (OBS.\,B, CD, E, 2001 and 2007) to the best 
fit continuum models are shown in Figure~\ref{MOS}. Several absorption profiles are 
clearly visible (which are labelled on each of the corresponding spectral panels), 
and they are generally in good agreement with the features present in the RGS in 
Figure~4. In particular, the broad absorption profiles present during OBS.\,CD and 
OBS.\,E are confirmed at high significance, whereby both the broad profile at 
$\sim1.2$\, keV (OBS.\,CD) and the complex absorption structure between 0.9--1.2\,keV 
(OBS.\,E) emerge at high signal to noise in the MOS data. Likewise the broad absorption trough 
detected at 1.06\,keV in the 2001 dataset, is also confirmed in the MOS spectra, with self 
consistent parameters. On the other hand, the featureless nature 
of the 2007 spectrum is also confirmed in the MOS, with no obvious residuals 
present in the spectrum. The results of the Gaussian line fitting to the MOS spectra are 
subsequently summarized in Table 3.

As a further check of the consistency between the RGS and MOS detections of the 
absorption lines, we generated and overlaid confidence contours for each of the lines that are  
independently detected in each of the RGS and MOS spectra. As an illustration we show two 
of the examples from the OBS.\,E and 2001 sequences in Figure\,7. The upper panel shows the 68\%, 
90\% and 99\% confidence contours (for 2 parameters of interest) of line energy against flux 
for the $\sim1.16$\,keV absorption line detected in the OBS.\,E spectra. The contours show the 
close agreement between the two detectors, with the absorption line confirmed in both the RGS and 
MOS spectra. Generally the line fluxes are more tightly constrained in the MOS, 
which is likely due to the higher signal to noise and 
better continuum determination afforded by the 
MOS spectra. The lower panel of Figure\,7 shows the comparison between line energy and width 
(in terms of $\sigma$) for the $\sim 1$\,keV broad absorption trough in the 2001 observation. 
The contours show that the broadening of the line is independently measured by both RGS and MOS, 
with a width of $\sigma\sim40$\,eV (or $\sigma=12000$\,km\,s$^{-1}$). The only marginal disagreement 
at the 90\% level is in the line energy, which is offset by $\sim20$\,eV, although this is likely to lie 
within the uncertainty of the absolute energy scale of the MOS detectors. 

\section{Modeling with Photoionized Absorption Models}

Having established the consistent detection of the absorption lines between the RGS and MOS 
spectra, in this next section we model the absorption (and any emission) from these spectra 
with a self consistent {\sc xstar} photoionization model (Kallman \& Bautista 2001). 
The goals of the modeling are to determine the likely identifications of the absorption 
profiles and to reveal the properties and possible outflow velocities of the soft X-ray absorption 
zones.

\subsection{Generation of Tabulated Photoionized Models}

In order to generate the \textsc{xstar} grids for the modeling, 
we used the input SED defined from a simultaneous fit to the 
\xmm\ OM, EPIC and {\it NuSTAR} data, taken from the 2013--2014 campaign. The details 
of the SED modeling can be found in Nardini et al. (2015); in summary the SED can be 
parameterized by a double broken power law from between the optical--UV, UV to soft 
X-ray and X-ray portions of the spectrum. In the optical-UV band covered by the 
OM (2--10\,eV), the spectrum has $\Gamma \simeq 0.7$ (so rising in $\nu F_{\nu}$ space), 
between the UV and soft X-rays (10--500\,eV) the SED declines steeply with $\Gamma \sim 
3.3$, and above 500\,eV, the intrinsic photon index is $\Gamma \simeq 2.4$. As per 
Nardini et al. (2015), the total 1--1000\,Rydberg ionizing luminosity is 
$5\times10^{46} -1\times10^{47}$\,erg\,s$^{-1}$, depending on the exact SED model 
adopted. Note that the main impact upon using this SED form as the input photoionizing 
continuum is in the ionization balance of the gas; when compared to a model computed 
with a standard $\Gamma=2$ X-ray power-law continuum, the ionization parameter will 
be comparatively higher, as there are a larger number of input UV and soft X-ray 
ionizing photons in the SED form, resulting in a higher ionizing luminosity over 
the 1--1000\,Rydberg band. This is particularly critical in the case of PDS\,456, 
which has a steep UV to X-ray continuum; as a result, the ionization parameter can 
be an order of magnitude higher when compared to a model produced with a flat 
$\Gamma=2$ illuminating continuum.

Subsequently, a tabulated grid of \textsc{xstar} models was calculated adopting 
this double broken power-law SED, with an input ionizing luminosity of 
$1\times10^{47}$\,erg\,s$^{-1}$ and an electron density of $n_{\rm e}=10^{8}$\,cm$^{-3}$, 
although the output spectra are largely insensitive to the density that is chosen. 
The grid of models also covers a wide range of parameter space, from 
$5\times10^{20} < N_{\rm H} <2.5\times10^{23}$\,cm$^{-2}$ in column density 
(in steps of $(\log N_{\rm H})=0.14$) and 
from $2 < \log\xi < 7$ (in steps of $(\log\xi)=0.2$) 
in terms of the ionization parameter. The subsequent output 
spectra were calculated over the 0.1--20\,keV range, over 10,000 energy steps. A 
turbulence velocity of $\sigma_{\rm turb}=5,000$\,km\,s$^{-1}$ was chosen for the 
final grid of models, as it was found that much lower turbulences could not match 
the breadth or equivalent width of the soft X-ray absorption features, while for 
higher turbulences, the absorption profiles became too broad to model the observed 
features seen near to 1\,keV in the RGS spectra.

Given the good agreement between the RGS and MOS for the \xmm\ sequences, for each 
sequence they were fitted simultaneously, allowing for a cross normalization factor 
(typically within $\pm5$\% of 1.0)
between the two instruments. This has the benefit of utilizing both the high spectral 
resolution of the RGS and the higher signal to noise of the MOS data. The spectra were 
fitted between 0.6--2.0\,keV, avoiding the energy band below 0.6\,keV in order to 
minimize any calibration uncertainties associated with the O edge in both detectors. 
We did not directly fit the MOS spectra above 2\,keV, as our motivation was to study 
the properties of the soft X-ray absorption. However, as a consistency check we did 
extrapolate the spectra above 2\,keV to see whether, to first order, the model reproduced 
the continuum form above 2\,keV. The continuum model adopted was the same as in the 
Gaussian fitting, i.e. a powerlaw plus blackbody component absorbed by Galactic 
absorption. Generally, the photon index of the power law was fixed in all of the 
sequences to $\Gamma=2.1$, however in OBS.\,B and 2007 this was found to be steeper 
with $\Gamma=2.4-2.5$. In the 2001 sequence, which has a strongly absorbed spectrum, 
an additional neutral partial covering absorption component (with $N_{\rm H}=4.8^{+1.1}_{-0.9} 
\times 10^{22}$\,cm$^{-2}$ and covering fraction $f_{\rm cov}=0.34\pm0.02$) was applied 
to the soft X-ray power law, as otherwise the photon index was found to be unusually 
hard with $\Gamma\sim1$. We note that a similar partial covering component has been 
found in some of the previous {\it Suzaku} observations of PDS\,456, when the spectrum 
becomes unusually hard (Reeves et al. 2014, Gofford et al. 2014, Matzeu et al. 2016).

\subsection{Fitting Methodology}

The \textsc{xstar} absorption grids were then applied to each of the sequences 
allowing the column density, ionization parameter and outflow velocity to vary 
independently, as well as the continuum parameters. 
In order to find the best 
fit solution for each absorption zone as fitted to each sequence in velocity space, we first varied the 
outflow velocity from $-0.3\,c$ (where the negative sign here denotes blueshift) 
up to $+0.3c$, in steps of $0.005c$, while 
allowing the ionization and column to adjust at each step in the fitting. 
Thus no a priori assumption is made about the likely velocity of any of the 
absorbers fitted to each sequence and the wide range in velocity space searched 
ensures that the global minimum is found, allowing for any possible zero velocity 
solution as well as outflow or inflow. 
This procedure was then repeated for each absorption zone which was subsequently added 
to each sequence, applying the criteria 
that the addition of each absorption grid to the 
model required an improvement in fit statistic corresponding to at least the 99.9\% 
confidence level. Once the global minimum for each absorption grid was found, we then 
conducted a finer search in velocity, ionization and column density in order to find the 
exact minimum and uncertainties associated with each parameter.

As an example, we describe in detail how this was applied to the OBS.\,E spectra, 
while the same fitting procedure was applied to all of the sequences. 
The initial starting point was with the baseline continuum model, 
with only the Galactic absorption component 
present (and no intrinsic absorption), which yielded a poor fit statistic of 
$\chi^{2}/\nu=289.0/161$. A single ionized zone of absorption 
was then added to the model and 
the $\chi^{2}$ space searched by stepping through the fit at each point in the 
outflow velocity, as described above. Figure 8 (upper panel) shows the resulting 
fit statistic against outflow velocity for this absorber. A clear minimum is found at 
$v=-0.17\pm0.01c$, while a zero velocity solution is ruled out at $>99.99$\% 
confidence. As a result of the addition of this single absorber with an outflow 
velocity of $-0.17c$, the fit statistic improved to $\chi^{2}/\nu=206/158$.
The corresponding F-test null hypothesis upon adding the first zone of absorption 
to the model (compared to the model with no absorption) 
is very low, with $P_{\rm f}=1.3\times10^{-11}$.
A second absorption grid was then added to the model, to test whether the fit improved 
further upon its addition and the velocity space of the new absorber was searched for any 
minimum in $\chi^{2}$. 
The result of adding this second absorption grid to the model is shown in Figure\,8 
(lower panel), which also shows a well defined minimum, but 
at a lower outflow velocity of 
$v=-0.06\pm0.01c$. The fit statistic also significantly improved upon the addition of 
the second absorption grid, with a resulting reduced chi-squared of 
$\chi^{2}/\nu=181/155$. 
The corresponding F-test null hypothesis upon adding the second zone of absorption 
to the model (compared to the model with only one absorber) 
is then $P_{\rm f}=1.6\times10^{-4}$. 
The addition of further absorption grids to the OBS.\,E spectra did not significantly improve the fit.

Thus the OBS.\,E spectrum requires two absorption zones, one with a faster 
outflow velocity ($-0.17c$), 
while both zones formally require an outflow velocity at $>99.9$\% 
confidence. 
The resulting best-fit absorption parameters for both OBS.\,E as well as the 
other spectral sequences are shown in Table\,4, after applying 
the same fitting procedure as described above. 
In general, 
the soft X-ray absorption can primarily be modeled with two main \textsc{xstar} 
zones (labeled as zones 1 and 2 respectively in Table 4), 
both of which are required to be outflowing with respect to the systemic velocity 
of PDS\,456, while we refer to zone 2 as having the higher velocity of the two zones. 

\subsection{Results of Photoionization Modeling to OBS.\,E}

\subsubsection{A Self-consistent Emission and Absorption Model}

In addition to the two absorption zones described above, the fit 
to OBS.\,E improves further to $\chi^{2}/\nu = 143.2/150$, by adding 
two broad Gaussian emission profiles. The emission line centroids are at 
$922\pm9$\,eV and $1066\pm16$\,eV and are apparent in the MOS and RGS spectra 
for OBS.\,E (see Figure 9), while the lines have fluxes of 
$7.5^{+4.0}_{-2.9}\times10^{-5}$ and 
$3.2^{+2.2}_{-1.8}\times10^{-5}$\,photons\,cm$^{-2}$\,s$^{-1}$, respectively 
(or equivalent widths of $17\pm8$\,eV and $11\pm7$\,eV). Note that strong 
0.92\,keV line emission has also been detected previously in some of the 
\suzaku\ XIS spectra of PDS\,456, see for instance Figure\,2 in Reeves et al. 
(2009). The line width, assuming a common width between the lines, is 
$\sigma=20\pm8$\,eV (or equivalently, $\sigma_{\rm v}=6,500\pm2,600$\,km\,s$^{-1}$). 

In order to model the emission, we removed the adhoc Gaussian emission lines from the model 
and instead fitted the emission using a self consistent grid of photoionized emission 
model spectra produced by \textsc{xstar}, with the same input parameters and 
turbulence velocity ($\sigma_{\rm turb}=5,000$\,km\,s$^{-1}$) as per the absorption 
grids. Like for the absorption modeling, two emission grids were added to the model, 
with the column densities of the two emission zones tied to the corresponding values 
in the two absorption zones. The ionization, outflow velocity and normalization 
(which is proportional to luminosity) of the two emission grids were allowed to vary. 
The parameters of the two emission zones are reported in Table\,5 as applied 
to OBS.\,E. The most significant emission zone is zone\,1, which is also the lower 
ionization of the two zones ($\log\xi = 2.8\pm0.3$) and yields an improvement in 
fit statistic of $\Delta \chi^{2}/\Delta\nu = 23.0/3$ upon its addition to the 
baseline absorption model, while in comparison the second and more highly ionized 
zone ($\log\xi\sim4.6$) is only marginally significant 
($\Delta \chi^{2}/\Delta\nu = 12.0/3$).  

This final \textsc{xstar} model provides a good fit to the simultaneous RGS and MOS 
spectra in OBS.\,E, with an overall fit statistic of $\chi^{2}/\nu = 146.5/149$. 
The resulting fit is shown in Figure 9, panel (a), with the \textsc{xstar} model 
able to reproduce the absorption features, i.e., the two broad absorption troughs 
at rest frame  energies of $\sim1.0$\,keV and $\sim1.2$\,keV. 
While the emitters do not necessarily have to be physically associated to the 
corresponding absorption zones, 
the emission is nonetheless able to reproduce the excess soft X-ray emission. In 
particular, the lower ionization emission zone 1 accounts for the stronger emission 
line observed near 920\,eV. Emission zone 1 also predicts an emission feature near 
to 0.8\,keV, which is consistent with the spectrum shown in Figure\,9. Emission zone 
2 has a more subtle effect, mainly accounting for the weaker emission line near 
1.06\,keV. 
If this best-fit model to the soft X-ray spectrum of OBS.\,E is extrapolated up to 
higher energies using the MOS data, the agreement is good, with the model reproducing 
well the shape and level of the continuum above 2\,keV, as is seen in Figure 9, panel 
(b). The remaining residuals at high energies are due to the iron K absorption profile, 
as reported in Nardini et al. (2015).

\subsubsection{Properties of the Soft X-ray Emission and Absorption}

In terms of the absorber properties, zone 1 has a lower column and lower ionization 
compared to zone 2 (see Table 4 for parameter details), as well as a lower outflow velocity 
of $v_{\rm out}/c=-0.064\pm0.012$ (compared to $v_{\rm out}/c=-0.17\pm0.01$ in zone 2), 
and primarily accounts 
for the lowest energy of the two absorption troughs. 
Note that a non-zero outflow velocity 
is required for both zones in the fit, as is also described in the fitting 
procedure described in Section\,5.2 (see Figure 8).
With the emitter included, if the velocity of both zones is forced to 
zero and the model refitted allowing the other absorber parameters to vary 
($N_{\rm H}$, $\log\xi$), then the fit is substantially worse compared to the case 
where the outflow velocities are allowed to vary ($\chi^{2}/\nu = 181.2/152$ versus 
$\chi^{2}/\nu = 143.2/150$). On the other hand, if we do allow the velocity to vary, 
but assume that it is the same for both zones, then the fit statistic is also worse 
than before ($\chi^{2}/\nu = 176.4/151$), as adjusting the ionization alone is not 
enough to explain the different energy centroids of the absorption profiles. 
Thus the two absorber zones are required to be distinct in velocity space, as well as 
in column/ionization.

In Figure 10 the relative contribution of each absorption zone is shown, in terms 
of the fraction of the continuum that is absorbed against energy. The main contribution 
towards zone 1 arises from Ne\,\textsc{ix--x}, as well as absorption from L-shell Fe from 
ions in the range Fe\,\textsc{xix--xx}; the blend of these lines together with a modest 
net blueshift produces the overall broad absorption trough which is centered near to 
1\,keV in the rest frame. The contribution towards zone 2 is similar, giving a blend 
of lines from Ne\,\textsc{ix--x} and Fe\,\textsc{xx--xiv}; then the higher energy of 
this absorption blend (centered near to 1.2\,keV), arises primarily from the higher 
outflow velocity and somewhat from the higher ionization of this zone. 

Note that although either the 1 keV or 1.2 keV absorption troughs in OBS.\,E
could at first sight be associated with absorption from Ne\,\textsc{x} (at 1.02\,keV)  
or Fe\,\textsc{xxiv} (at 1.16\,keV) with a smaller blue-shift, the self consistent 
{\sc xstar} modeling presented above appears to rule this out. 
This is because it is difficult 
to produce absorption from these species in isolation, without a contribution from 
other, lower energy, lines. 
For instance for the higher velocity zone, although the Fe\,\textsc{xxiv} $2s\rightarrow3p$ 
absorption can contribute towards the broad absorption profile at 1.2\,keV, 
the other lower energy L-shell Fe lines near to 1\,keV also strongly contribute, 
reducing the overall centroid energy of the absorption trough produced in the 
{\sc xstar} model (see Figure 10 for their relative contributions). 
The net result is that an overall blue-shift of the absorption profile is required 
to match its observed energy in the actual spectrum.

In terms of the emission, the strongest contribution is from 
zone 1, which has the lowest ionization parameter. The outflow velocities of the two 
emission zones are lower than their respective absorption zones, i.e., the outflow 
velocity of zone 1 is $v_{\rm out}=-0.04\pm0.02c$. This could be 
consistent with the emission 
expected from a bi-conical outflow, as emission can be observed across different 
sightlines which will appear to have a lower net outflow velocity than along the direct 
line of sight to the observer. The emission from the near side of the outflow 
may be preferentially observed in this case, if the emission from the far (receding) 
side of the wind is obscured through a greater column of matter. This may be the case 
if, for instance, the receding wind is on the far side of the accretion disk.

The relative contribution of the photoionized emission 
zones to the soft X-ray spectrum of PDS\,456 are shown in Figure 11. The dominant 
contribution towards the emission is from zone 1 (Figure 11, red curve) and the majority 
of the emission arises from O\,\textsc{vii} and O\,\textsc{viii}. In particular, the 
emission near to 0.9\,keV is mainly accounted for by blueshifted O\,\textsc{viii} 
Radiative Recombination Continuum (RRC) at 0.87\,keV, 
along with some contribution from the forbidden line of Ne\,\textsc{ix}. 
At first sight, the O\,\textsc{viii} RRC appears 
stronger than the O\,\textsc{viii} Ly$\alpha$ line at lower energies. However, this is 
just due to the latter emission being more suppressed by the Galactic absorption towards 
PDS\,456, as this lower energy line falls just above the deep neutral O\,\textsc{i} edge 
as seen in the observed frame. In contrast, the higher ionization ($\log\xi=4.6^{+0.6}_{-0.4}$) 
zone 2 is much weaker (Figure 11, blue curve), and its main contribution is to reinforce 
the emission near 1\,keV, mainly due to blueshifted emission from Ne\,\textsc{ix} and 
Ne\,\textsc{x}. 

\subsection{Extension to the other \textit{XMM--Newton} observations}

The absorption modeling was then extended to the other 2013--2014 sequences, 
as well as the 2001 and 2007 \xmm\ observations. In these observations, the soft 
X-ray emission is less prominent, as the continuum fluxes are generally higher 
than in OBS.\,E, thus we subsequently concentrated on modeling only the absorption 
zones. The zone 2 absorber appears common to all of the data sets, except for the 
bright and featureless 2007 spectrum, where only an upper limit can be placed on 
the column. Indeed, the ionization parameter of this zone is consistent within 
errors for all the data sets, varying only in the narrow range from 
$\log\xi=4.04-4.19$ (see Table\,4). The main differences arise in the column density, 
which is higher in the more absorbed OBS.\,E ($N_{\rm H}=1.5\pm0.4 \times 10^{22}$\,cm$^{-2}$) 
and 2001 ($N_{\rm H}=2.2^{+0.7}_{-0.6} \times 10^{22}$\,cm$^{-2}$) sequences and weakest 
in the least absorbed OBS.\,B and 2007 observations. Indeed in the bright, continuum dominated 2007 observations, 
a low upper-limit can be placed on the column density of $N_{\rm H}< 0.23 
\times 10^{22}$\,cm$^{-2}$. Likewise, although the RGS data is not available 
for OBS.\,A, on the basis of the EPIC/MOS data we can still place an upper-limit on the 
column density of $N_{\rm H}<0.35\times10^{22}$\,cm$^{-2}$ for zone 2, consistent with the absorption being 
weaker while the continuum flux was brighter during this observation. 

The outflow velocity is strongly variable between some of sequences, e.g. 
$v_{\rm out}/c=-0.27\pm0.01$ in OBS.\,CD versus $v_{\rm out}/c=-0.17\pm0.01$ in 
OBS.\,E. This is illustrated in Figure 12, which shows the confidence contours 
in the $N_{\rm H}$ versus outflow velocity plane for zone 2 in each of the 
B, CD, E and 2001 sequences, the best fit parameters clearly differing along 
the velocity axis. Thus, while the column determines the depths of the broad 
absorption troughs, the outflow velocity determines its blueshift. Indeed, in 
the CD observation it can be seen (e.g. from Figure 6) that the broad absorption 
trough is shallower and shifted to higher energies compared to observation E. 

On the other hand, while the lower velocity zone 1 is required to reproduce 
the lower energy absorption troughs in OBS.\,E and 2001 near to 1\,keV, it is 
not required in any of the other observations and thus appears rather sporadic. 
Zone 3 is included for completeness and represents a possible high ionization 
phase of the wind, similar to what is seen at iron K, with a similar ionization 
level of $\log \xi = 5.5$
and a velocity of $-0.3\,c$ (i.e., 
Nardini et al. 2015). At this level of ionization, in the soft X-ray band, there 
is only a small imprint from zone 3 in the spectrum, which is mainly in the form 
of a O\,\textsc{viii} Ly$\alpha$ absorption line blueshifted to above 0.8\,keV 
(see the lower panel of Figure 10), while only trace amounts of absorption due 
to H-like Ne and Mg are predicted. Due to its transparency at soft X-rays, this 
highest ionization zone is therefore only formally required in the OBS.\,CD spectrum, 
mainly to reproduce the weak absorption feature detected near to 0.84\,keV in the 
RGS and MOS data (see Tables\,2 and 3). 

\section{Discussion}
\subsection{The Detection of Fast Soft X-ray Absorbers}

Although the numbers of confirmed fast outflows have significantly increased 
through systematic studies of AGN in the iron K band (Tombesi et al. 2010, Gofford 
et al. 2013), to date there have been relatively few detections of fast outflows in 
the soft X-ray band. Here, the detection of broad soft X-ray absorption line profiles 
associated with the fast wind in PDS\,456 represents one of the few cases where the 
presence of an ultra fast outflow has been established in the form of  highly 
blueshifted absorption in both the Fe K and soft X-rays bands. Other notable recent 
examples include the nearby QSO, PG\,1211+143 (Pounds 2014) and in the NLS1, 
IRAS\,17020+4544 (Longinotti et al. 2015). In both of these cases, 
the observations suggest that the outflow is more complex than a single-zone medium, 
with multiple ionization and/or velocity components. In PG\,1211+143, at least two 
different velocity components are present, both in the iron K band as well as soft 
X-rays (Pounds et al. 2016), while in the NLS1 IRAS\,17020+4544, a complex fast 
($v_{\rm out}=23,000-33,000$\,km\,s$^{-1}$) soft X-ray absorber is present, covering 
at least 3 zones in terms of ionization parameter. 

In PDS\,456, while the velocity of the soft X-ray absorber (with $v_{\rm out}=0.17-0.27\,c$ 
for zone 2) is similar to the highly ionized iron K absorption in this AGN 
($v_{\rm out}=0.25-0.3\,c$), the column density and ionization is up to two orders 
of magnitude lower than the iron K absorber, i.e., $\log\xi \sim4$ and 
$N_{\rm H}=10^{22}$\,cm$^{-2}$ at soft X-rays versus $\log \xi \sim 6$ and 
$N_{\rm H}=10^{24}$\,cm$^{-2}$ at iron K. This also suggests that the outflow 
is more complex than what would be expected for a simple homogeneous radial outflow, 
where the density varies with radius as $n\propto r^{-2}$. Thus, as $\xi=L/nr^{2}$, 
in the radial case the ionization should remain approximately constant along the outflow. 
%Thus either the outflow geometry or structure is likely more complex. 
A density profile with a power-law distribution flatter than $n(r) \propto r^{-2}$ 
could account for a decrease in ionization with absorber radial distance, 
and indeed it has been suggested from the distribution of warm absorbing 
gas amongst several Seyfert 1s (Behar 2009, Tombesi et al. 2013). Another 
likely possibility, especially given the rapid variability of the soft X-ray 
absorption seen here, is a clumpy, multi phase wind. Thus denser (and more 
compact) clumps or filaments could coexist within a smoother highly ionized 
outflow to explain the ionization gradients. Indeed a recent, albeit slower 
($v\sim0.01c$), multi-phase disk wind was recently revealed in the changing 
look Seyfert galaxy, NGC\,1365 (Braito et al. 2014), with very rapid changes 
in soft X-ray absorption (with $\Delta N_{\rm H}\sim10^{23}$\,cm$^{-2}$ in 
$\Delta t <100$\,ks), requiring a clumpy outflowing medium.

\subsection{The Soft X-ray Outflow}

We now explore the properties of the soft X-ray absorber within the context 
of a clumpy wind model. In order to place a radial constraint on the absorbing gas, 
its variability timescale is considered. We concentrate on zone 2, which is the best 
determined of the absorption zones and is relatively constant in ionization 
(see Table\,4). During the first four observations in the 2013--2014 campaign, 
there is a subtle variation in the column density from $<0.35\times10^{22}$\,cm$^{-2}$ 
(OBS.\,A) to $0.70^{+0.22}_{-0.21}\times10^{22}$\,cm$^{-2}$ (OBS.\,CD), i.e. a 
slight increase over a 3 week timescale. A stronger variation occurs over the 
subsequent 5-month period between OBS.\,A--D and OBS.\,E, with an increase in 
column density to $1.5\pm0.4\times10^{22}$\,cm$^{-2}$, while the velocity of the 
absorber also decreases from $0.27\pm0.01\,c$ to $0.17\pm0.01\,c$.
The more subtle absorption variability between OBS.\,A and OBS.\,D in 2013 
likely corresponds to the passage of the same absorbing system (with the same 
outflow velocity) across the line of sight. However, the stronger variations 
in $N_{\rm H}$ and outflow velocity between this and OBS.\,E (as well as in 2001) 
likely correspond to different absorption systems (with 2007 being unabsorbed).
Therefore we set a plausible timescale for the absorption variability, in terms 
of the passage of clumps or filaments of gas across our line of sight, of between 
$\Delta t = 10^{6} - 10^{7}$\,s, i.e., weeks to months. It is also likely that 
some variability can occur on shorter timescales, as found during the low flux 2013 
{\it Suzaku} observations, where changes in the soft X-ray spectrum on timescales 
of $\gs100$\,ks could be accounted for by variability of a partially covering 
absorber (Matzeu et al. 2016). 

The transverse motion of the absorbing systems moving across the line of sight 
is assumed to be due to the overall Keplerian rotation of the gas within the wind, 
with the observed outflow velocity being due to the bulk motion of the wind towards 
us. The Keplerian velocity at a radius $R$ is simply $v_{\rm K}^{2}=c^{2}/r_{\rm g}$, 
where $r_{\rm g} = R / R_{\rm g}$ is the radial distance from the black hole in 
gravitational units. The column through an (approximately spherical) absorbing 
cloud can be expressed as $N_{\rm H} \sim n_{\rm H} \Delta R$, where its size is 
$\Delta R = v_{\rm K} \Delta t$. The number density of an absorbing cloud is then 
given as:
\begin{equation}
n_{\rm H} \sim \frac{N_{\rm H}}{\Delta R} = \frac{N_{\rm H} r_{\rm g}^{1/2}}{c\Delta t}.
\end{equation}
The definition of the ionization parameter gives $n_{\rm H} \sim n_{\rm e} = 
L_{\rm ion}/\xi R^{2}$, thus equating these densities yields the radial distance 
of the clouds:
\begin{equation}
r_{\rm g}^{5/2} = \frac{L_{\rm ion}}{\xi} \frac{c^{5}\Delta t}{N_{\rm H}} (GM_{\rm BH})^{-2}.
\end{equation}
For the zone 2 absorber $\log\xi \sim 4$ and $N_{\rm H} \sim 10^{22}$\,cm$^{-2}$, 
while for PDS\,456 $L_{\rm ion} \sim 10^{47}$\,erg\,s$^{-1}$ and $M_{\rm BH}\sim10^{9}\Msun$. 
Then for $\Delta t = 10^{6}-10^{7}$\,s, $R \sim 4000-11000R_{\rm g} = 0.7-1.7 \times 10^{18} 
\sim 10^{18}$\,cm. At this distance, and for the ionization of the gas, then the 
number density is $n_{\rm H}\sim10^{7}$\,cm$^{-3}$, while the radial extent of the 
clouds is $\Delta R\sim10^{15}$\,cm. So the outflow clouds can be fairly compact 
(approximately $10 R_{\rm g}$), and also capable of partially covering the X-ray 
source. 

At a distance of $R\sim10^{18}$\,cm, the Keplerian velocity is 
$v_{\rm K}\sim4000$\,km\,s$^{-1}$. This agrees well with the measured velocity 
widths of the soft X-ray emission lines, with $\sigma_{\rm v}=6,500\pm2,600$\,km\,s$^{-1}$ 
(e.g. from OBS.\,E), or the required velocity broadening ($\sigma=5,000$\,km\,s$^{-1}$) 
in the \textsc{xstar} modeling. Furthermore, this matches well the velocity widths 
of the broad UV emission lines in PDS\,456, where the FWHM of the Ly$\alpha$ and 
C\,\textsc{iv} lines are 12,000\,km\,s$^{-1}$ and 15,000\,km\,s$^{-1}$, 
respectively, and where the C\,\textsc{iv} profile shows one of the largest 
blueshifts amongst quasars of $\sim5,000$\,km\,s$^{-1}$ (O'Brien et al. 2005; 
compare to Richards et al. 2002).
Thus the location of the soft X-ray outflow appears coincident with the BLR in 
this AGN, and indeed some of the more ionized (and less dense) BLR clouds could be responsible for 
the soft X-ray absorption. 

Alternatively, we can assume instead that the soft X-ray absorber arises from a 
wind with a smooth radial profile out to large distances. In that case, from integrating 
through the flow, then the radial extent of the wind is given by $R_{\rm smooth} = 
L_{\rm ion} / N_{\rm H} \xi \sim 10^{21}$\,cm for the above outflow parameters. Such 
a large-scale outflow up to a kpc in extent, with a rather low density (where 
$n_{\rm H}\sim10$\,cm$^{-3}$), would not be able to explain the variability of 
the absorber on weeks to months timescales. Furthermore, the velocity widths 
resulting from gas on these large scales is likely to be rather small, of 
$\sigma \sim 100$\,km\,s$^{-1}$, whereas (as seen in Figure\,2) there is no 
narrow component of emission/absorption in PDS\,456 that could be associated to 
a distant warm absorber. We conclude that the fast, soft X-ray absorber in 
PDS\,456 arises from clumpy gas on smaller (BLR) scales, and may exist 
co-spatially as a higher density (but lower ionization) phase of the fast, 
accretion disk wind. 

\subsection{The Soft X-ray Emission}

The luminosity of the soft X-ray line emission can also be used to calculate 
the global covering factor of the outflowing gas. From the photoionization modeling, 
the normalization (or flux), $\kappa$, of each of the emission components is defined 
by {\sc xstar} in terms of:
\begin{equation}
\kappa = f\frac{L_{38}}{D_{\rm kpc}^2}
\end{equation}
where $L_{38}$ is the ionizing luminosity in units of $10^{38}$\,erg\,s$^{-1}$, 
$D_{\rm kpc}$ is the distance to the quasar in kpc. Here $f$ is the covering fraction 
of the gas with respect to the total solid angle, where $f = \Omega / 4\pi$. For a 
spherical shell of gas, $f=1$, while $L$ is the quasar luminosity that illuminates 
the photoionized shell. Thus by comparing the predicted normalisation ($\kappa$) for 
a fully covering shell of gas illuminated by a luminosity $L$ versus the observed 
normalization ($\kappa_{\rm xstar}$) determined from the photoionization modeling, 
the covering fraction of the gas can be estimated. For PDS\,456, with 
$L=10^{47}$\,erg\,s$^{-1}$ at a luminosity distance of $D=860$\,Mpc, for a spherical shell 
the expected {\sc xstar} normalization is $\kappa=1.35\times10^{-3}$. Compared to 
the observed normalization factors reported in Table\,5, the covering fraction of 
strongest lower ionization (zone\,1) of emitting gas is $f=0.44\pm0.22$. 

The soft X-ray emitting (and absorbing) gas thus covers a substantial fraction of 
$4\pi$ steradian, consistent with the result obtained by Nardini et al. (2015) 
for the high ionization wind measured at Fe K, which was found to cover at least 
$2\pi$\,steradian solid angle. This is consistent with the picture of the soft X-ray 
outflowing gas being embedded within the wide angle high ionization disk wind. Note 
that, despite the large covering factor of the soft X-ray gas, its {\it volume} 
filling factor is likely smaller. This is consistent with a geometry where the 
absorbing clouds intercept a large enough fraction of the sightlines from the X-ray 
source, but individually are compact enough to occupy a small enough region in 
volume.

Indeed, the volume filling factor can be estimated by considering the emission 
measure of the gas. To estimate the emission measure, we consider the strong 
O\,\textsc{viii} RRC emission, with a flux of 
$\sim7.5\times10^{-5}$\,photons\,cm$^{-2}$\,s$^{-1}$, as determined from the 
fit to the \xmm\ OBS.\,E spectrum (see Section 5.2). This corresponds to a 
luminosity of $L_{\rm O\,VIII} \sim 6\times10^{51}$\,photons\,s$^{-1}$ at the 
distance of PDS\,456. The recombination coefficient for O\,\textsc{viii} is 
$\alpha_{r}=1.2\times10^{-11}$\,cm$^{3}$\,s$^{-1}$, for a temperature of 
$kT\sim10$\,eV (Verner \& Ferland 1996). The overall emission measure can then 
be calculated from:
\begin{equation}
{\rm EM} = \frac{L_{\rm O\,VIII}}{\alpha_r A_{\rm O}f_if_r}
\end{equation}
where $A_{\rm O}$ is the abundance of Oxygen, $f_i$ is the ionic fraction of the 
parent ion (fully ionized Oxygen) and $f_r$ is the fraction of recombinations 
that occur direct to the ground state. Here we take $f_i\sim0.5$ given the ionization 
state of the gas, $f_r\sim0.3$ for the fraction of recombinations direct to ground, 
while an O abundance of $4.9\times10^{-4}$ is assumed (Wilms et al. 2000, Asplund 
et al. 2009). This gives an estimated emission measure for the soft X-ray emitting 
gas of ${\rm EM}\sim6\times10^{66}$\,cm$^{-3}$.

In comparison we can also calculate the emission measure by assuming the gas clouds 
occupy a spherical region of radius $R$, with a mean density $n$ and a volume filling 
factor of $V_{\rm f}$. For the emission measure:-
\begin{equation}
{\rm EM} = \int n^2 {\rm d}V \sim \frac{4}{3}\pi R^{3} V_{\rm f} n^{2}.
\end{equation}
Taking a radius of $R=10^{18}$\,cm (consistent with the line velocity widths) and an 
emitter ionization of $\log\xi\sim3$ (for the lower ionization zone 1) gives a density 
of $n=10^8$\,cm$^{-3}$ for the emitting clumps. By comparison of the above expression 
with the emission measure estimated from the line emission, this implies a volume filling 
factor of $V_{f}\sim10^{-4}$, consistent with our expectations of a clumpy medium. A 
higher fraction of $V_{\rm f}\sim1$ only occurs at much larger kpc scales (and at 
correspondingly lower densities, as $nR^{2}=L_{\rm ion}/\xi$), resulting in a smooth 
medium, but inconsistent with the observed absorption variability. Note that the clumps 
probably exist as quasi-spherical clouds, rather than long strands or filaments of 
material, in order to satisfy the criteria of a low volume filling factor but a high 
covering factor.

\subsection{Outflow Energetics}
\noindent Following Nardini et al. (2015), we also estimate the mass outflow rate from:
\begin{equation}
\dot{M}_{\rm out}=1.2m_{\rm p}\times 4\pi f v_{\rm out} N_{\rm H} R
\end{equation}
where $f$ is the covering fraction and for the outflow, $v_{\rm out}=0.2c$, 
$N_{\rm H}=10^{22}$\,cm$^{-2}$, $R=10^{18}$\,cm and $f\sim0.4$ from above. As the 
above equation is expressed in terms of the column density, it is independent of 
any volume filling factor, as the observed $N_{\rm H}$ through the line of sight 
is the same regardless of whether the outflowing matter is smooth or clumpy. This 
yields a mass outflow rate of $\dot{M}_{\rm out}=6\times10^{26}$\,g\,s$^{-1}$ (or 
$\sim 10 M_{\odot}$\,yr$^{-1}$), with a corresponding kinetic power of 
$\dot{E}_{K}\sim10^{46}$\,erg\,s$^{-1}$ (or approximately 10\% of the bolometric 
luminosity). This is consistent with the estimates from Nardini et al. (2015) for 
the highly ionized zone of the wind. This could suggest either that the lower 
ionization clumps form further out along the wind from the high ionization matter, 
or that the less ionized matter carries an energetically similar component as per 
the highly ionized part of the disk wind.

\subsection{The overall view of the wind in PDS 456}
 
PDS\,456 is the first known AGN where it has been possible to resolve broad absorption 
profiles (with typical velocity widths of $\sigma\sim5,000$\,km\,s$^{-1}$) in the soft 
X-ray band. In contrast, the absorption lines detected from the few fast outflows that 
appear to be present at soft X-rays (e.g. Gupta et al. 2013, 2015, Longinotti et al. 2015, Pounds et al. 2016), 
appear to result from narrower systems. Indeed, the detection of both the broad soft 
X-ray absorption lines and the fast P-Cygni-like profile at iron K may suggest that 
PDS\,456 is a higher ionization, X-ray analogue of the Broad Absorption Line (BAL) 
quasars commonly known to occur in the UV (Turnshek et al. 1988, Weymann et al. 1991).
Indeed fast X-ray counterparts of the UV outflows have also been detected in 
some of the BAL quasars (Chartas et al. 2002, 2003). However unlike the UV outflows, 
which are likely to be line driven, the X-ray outflows are generally faster and higher ionization. 
Indeed they may instead be driven by other mechanisms, 
such as by continuum radiation, i.e. Compton scattering driven winds (King \& Pounds 2003), 
or by magneto hydrodynamical processes (Fukumura et al. 2010).

One possibility is that the broad soft X-ray absorption lines in PDS\,456 may become 
more apparent when the overall continuum is more absorbed. This appears to be the case 
in the current RGS observations, where the quasar spectrum is generally more featureless 
at higher flux levels, while the absorption structure becomes apparent when the continuum 
flux is attenuated. A similar situation may have occurred during the recent extended 
campaign on the nearby Seyfert 1, NGC 5548 (Kaastra et al. 2014), when broad and 
blueshifted UV BAL profiles unexpectedly emerged when the soft X-ray flux was heavily 
suppressed by a partial covering X-ray absorber. In PDS\,456, the X-ray spectrum is 
also highly variable and similar to NGC\,5548, as well as to other AGN such as NGC\,3516 
(Markowitz et al. 2008, Turner et al. 2008), NGC\,1365 (Risaliti et al. 2009) or 
ESO\,323-G077 (Miniutti et al. 2014), which exhibit pronounced absorption variability 
at soft X-rays. 

Indeed a prolonged absorption event may have occurred in PDS\,456 months prior to 
the 2013 \xmm\ observations, during an extended low flux observation with {\it Suzaku} 
in February--March 2013 (see Gofford et al. 2014; Matzeu et al. 2016 for details of 
these observations). During the 2013 {\it Suzaku} observations, the soft X-ray flux 
was lower by about a factor of $\times 10$, when compared to the relatively unobscured 
level observed in \xmm\ OBS.\,B just six months later (Figure\,13). This suggests the 
presence of a variable, high column density obscuring medium in PDS\,456, with 
$N_{\rm H}>10^{23}$\,cm$^{-2}$.  
Several months later, during the last of the current \xmm\ observations in February 
2014 (OBS.\,E), the soft X-ray flux below 2\,keV started to decline again when compared 
to OBS.\,B, coincident with when the broad soft X-ray absorption features emerged in 
the RGS spectra. The more moderate level of obscuration during OBS.\,E (with 
$N_{\rm H}\sim10^{22}$\,cm$^{-2}$), likely accounts for the decrease in soft X-ray 
flux between OBS.\,B and OBS.\,E, while the overall level of the hard X-ray continuum 
above 2\,keV remained unchanged.

The overall view of the wind in PDS\,456 thus appears more complex than a simple, 
homogeneous radial outflow. From the above considerations it appears that the soft 
X-ray outflow represents a denser, variable and clumpy absorber, likely embedded within 
the fast, less dense, higher ionization phase of the wide-angle wind responsible for 
the persistent iron K absorption profile. The fast, high ionization phase of the 
wind is likely launched from the inner accretion disk (at distances within 
$\sim100\,R_{\rm g}$ or $\sim10^{16}$\,cm from the black hole, see Nardini et al. 2015), 
while the clumps within the outflow appear to manifest over larger scales, at around 
$R\sim10^{18}$\,cm, and are perhaps coincident with density perturbations within the wind.
Indeed, hydrodynamical simulations naturally predict the existence of time variable 
density variations and streamlines within the wind (Proga, Stone \& Kallmann 2000), which 
could then lead to the observed X-ray absorption variability. The denser clumps responsible 
for the soft X-ray absorption appear radially coincident with the expected location of the 
AGN broad emission line region, while they may also be responsible for the soft X-ray 
re-emission from the wind. The denser clouds may also be sufficiently compact 
($\Delta R\sim10\,R_{\rm g}$) to act as the putative partial covering X-ray absorber, 
and may thus be responsible for the strong (order of magnitude) absorption variability 
seen at soft X-rays towards this quasar (Gofford et al. 2014, Matzeu et al. 2016). 
Indeed, in AGN in general, a complex topology of clouds surrounding the X-ray continuum 
source may be able to account for the general distribution of broadband X-ray spectral 
shapes observed from both the type I and type II AGN population (Tatum et al. 2013, 2016).

Future, high resolution observations may prove crucial in revealing the 
structure of the accretion disk wind seen in PDS\,456, as well as in other quasars, 
and the different wind phases that are present over different scales. 
Furthermore, any coincidence 
between the UV and the soft X-ray absorber in PDS\,456 can be established by simultaneous 
soft X-ray observations (i.e., with \xmm\ RGS or {\it Chandra} LETG) and the UV (via 
{\it HST}/COS), which, as for the case of NGC\,5548, can determine how the UV line 
profiles respond to changes in the soft X-ray absorber.

\section{Acknowledgements}

J.\,N. Reeves, E. Nardini, P.\,T. O'Brien and M.\,T. Costa acknowledge the financial support of STFC. 
J.\,N. Reeves also acknowledges NASA grant number NNX15AF12G, while T.\,J. Turner acknowledges
NASA grant number NNX13AM27G. E. Behar 
received funding from the European Unions Horizon 2020 research and innovation programme under the 
Marie Sklodowska-Curie grant agreement no. 655324, and from the I-CORE program of the Planning and 
Budgeting Committee (grant number 1937/12).  This research is based 
on observations obtained with \xmm, an ESA science mission with instruments and contributions 
directly funded by ESA Member States and NASA.

\section{Appendix: Statistical Significance of the Absorption Line Detections}

Here we quantify and discuss the statistical significance of the absorption line detections in the 
RGS and MOS spectra, which are reported in Section\,4 of the main article.
As an initial guide to the 
significance of the line detections, Tables 2 and 3 give the 
F-test null probabilities ($P_{\rm F}$) for the addition of each of the lines to the RGS and MOS spectra, 
noting that the F-test assumes that the prior distributions of the uncertainties are Gaussian. 
Nonetheless this gives an indication of the significance level. 
Considering the RGS alone, the most significant lines are the broad absorption profiles in 2001 
at 1.06\,keV and in the OBS.\,E spectrum at 1.16\,keV, where $P_{\rm F}= 1.3\times10^{-5}$ 
and $P_{\rm F}= 9.0\times10^{-6}$ respectively, which exceeds the $4\sigma$ level. 
The weakest two features (the emission 
line associated to OBS.\,E, and the weak 850\,eV absorption line in OBS.\,CD) 
are confirmed at about the 99\% level in the RGS alone.

However when considering the MOS spectra alone, where the statistics are highest, 
the null probabilities are even smaller, e.g. at the level of $P_{\rm F}\sim10^{-11}$ for some of the lines, 
which is at the $7\sigma$ confidence level. 
Indeed the likelihood of any false positives in both RGS and MOS is very small, 
as they would also have to occur at a consistent energy 
between both detectors.
For instance the above 1.16\,keV absorption line in OBS.\,E has null probabilities of 
$P_{\rm F}= 9.0\times10^{-6}$ (RGS) and $P_{\rm F}= 7.4\times10^{-6}$ (MOS) for each detector, 
and thus the resultant probability of a false detection occurring in both detectors at the same energy
is extremely small. 
This is also the case for the other RGS lines; indeed when considering their multiplicative 
probabilities of being independently detected in both the RGS and MOS, then the line detections are all 
subsequently confirmed at the $5\sigma$ or higher level.  

Indeed as a further guide to the whether the spectra can be adequately accounted for by 
a simple baseline continuum without lines (which is the null hypothesis case), 
we also quote the reduced chi-squared for the baseline model and subsequent false probability 
resulting from this in Tables 2 or 3. 
For each spectrum, the fit statistic is extremely poor 
without the addition of any absorption lines to the spectra and the baseline continuum 
alone is not an adequate description of the data.

Monte Carlo simulations have then been used to assess the significance of the line detections 
(Protassov et al. 2002). This is essentially a means to calibrate a test statistic, whereby the 
prior uncertainty distribution is not known.
As a test case, we ran the simulations on the OBS.\,E spectra, as this has the lowest continuum flux and 
thus may be more likely to yield a higher rate of false positives.  
The Monte Carlo simulations were performed independently for both the RGS and MOS exposures. 
For the null hypothesis model, we adopted the best fit baseline model for the continuum, 
with no absorption features, and we simulated  1000 spectra  with the photon statistics 
expected for  the same exposure times of the actual observation.  
In order to account for the uncertainties in the continuum model, each of the simulated spectra 
was then fitted with the null hypothesis model and a new simulation was performed with the 
new best fit as the continuum model; see Porquet et al. (2004) and Markowitz, Reeves \& Braito (2006) 
for similar examples of applying this method.

Each of these simulated spectra  was then  fitted  with the null hypothesis model  to obtain   a
baseline $\chi^2$  value, fitted over the full 0.4-2.0\,keV band. 
The spectra were systematically searched for any line-like deviations over the 
0.8--1.5\,keV (rest-frame) energy range, which is the most likely band for the lines to be found in the real data. 
To achieve this, we added a Gaussian line component and stepped through each energy interval in the spectra, with a step size of
7 eV and re-fitted the spectrum at each step. Thus by searching the spectra for any 
false positives over a wide energy range, no a priori assumption is made about the initial 
line energy and the Monte Carlo simulations represent a blind trial. 
Note that the step size over which the simulations were searched 
was chosen in order to  match the binning used in the actual RGS  and MOS spectra. 
We also assumed a line width of $\sim40$\,eV, similar to the typical value in the actual data.

We then  obtained for each  RGS  simulated spectrum a minimum $\chi ^{2}$ and  from these 
created a distribution of 1000 simulated values of the maximum $|\Delta \chi^{2}|$ values obtained from each spectrum (compared to the null hypothesis model). 
We then  constructed two independent cumulative frequency distributions
of the $|\Delta \chi^{2}|$ values expected for a blind line search in  both the MOS and RGS spectra. 

Only 1 RGS fake spectrum had a  $|\Delta \chi^{2}| \ge  17.3$, 
which corresponds to the lowest significance case of the reported 
$|\Delta\chi^2|$  values for the observed absorption lines in the 
actual OBS.\,E spectrum (at 1.02\,keV, see Table\,2). 
Thus the inferred rate of false detections is $P_{\rm F}=1\times10^{-3}$, 
or alternatively the  statistical significance of the detection is 99.9\%. 
We then inspected the $\Delta \chi^{2}$ distribution  of the MOS  simulated spectra 
and we found that none of the spectra has a $|\Delta \chi^{2}| \ge  33.4$, which is the level of the 
fit improvement in the actual MOS spectrum for the same 1\,keV absorption line. 
Indeed the highest  $|\Delta\chi^2|$ deviation found from the MOS simulations is  18.8;  
thus the false detection rate in the MOS OBS.\,E spectra is $P_{\rm F}<10^{-3}$.  Hence 
even without requiring the false detections  to occur at the same energy in the MOS and RGS
(i.e. within $\pm 20 $\,eV as above), the Monte Carlo simulations demonstrate 
that the absorption features in OBS\,E have a combined false probability of 
$P_{\rm F}<10^{-6}$, given that the false detections would have to occur in both the RGS and MOS. 
Furthermore, given the lines occur at the same energy in the actual RGS and MOS spectra, this 
confirms that the features are likely intrinsic to the AGN  
and are not a statistical artifact.

\clearpage

\begin{figure}
\begin{center}
\rotatebox{-90}{\includegraphics[width=11.5cm]{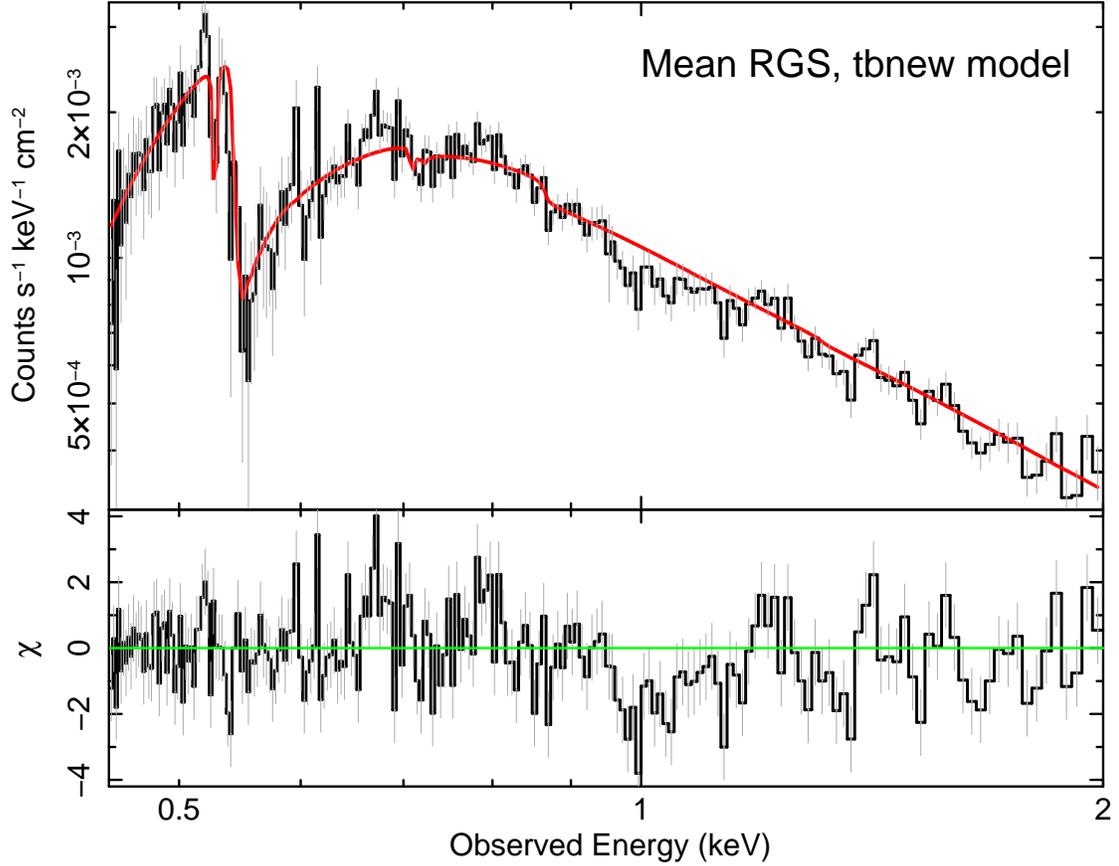}}
\end{center}
\caption{The summed RGS\,1+2 spectrum of PDS\,456, taken from the mean of 
the B--E 2013--2014 observations. The $1\sigma$ errors are shown in greyscale. 
The spectra have been rebinned with a constant resolution of $\Delta\lambda=0.1\,$\AA\ 
in wavelength and are plotted in the observed frame. This figure shows the RGS 
spectrum (upper panel) and residuals (in $\sigma$, lower panel) compared to the 
best-fit continuum model (solid red line) consisting of a power-law (of photon index 
$\Gamma\sim 2.1$) plus blackbody emission component. The Galactic absorption has 
been modeled with the \texttt{tbnew} ISM absorption model, as described in the text. 
Note the broad negative residuals present near 1\,keV in the observed frame}.
\label{tbnew}
\end{figure}

\clearpage

\begin{figure}
\begin{center}
\rotatebox{-90}{\includegraphics[height=14cm]{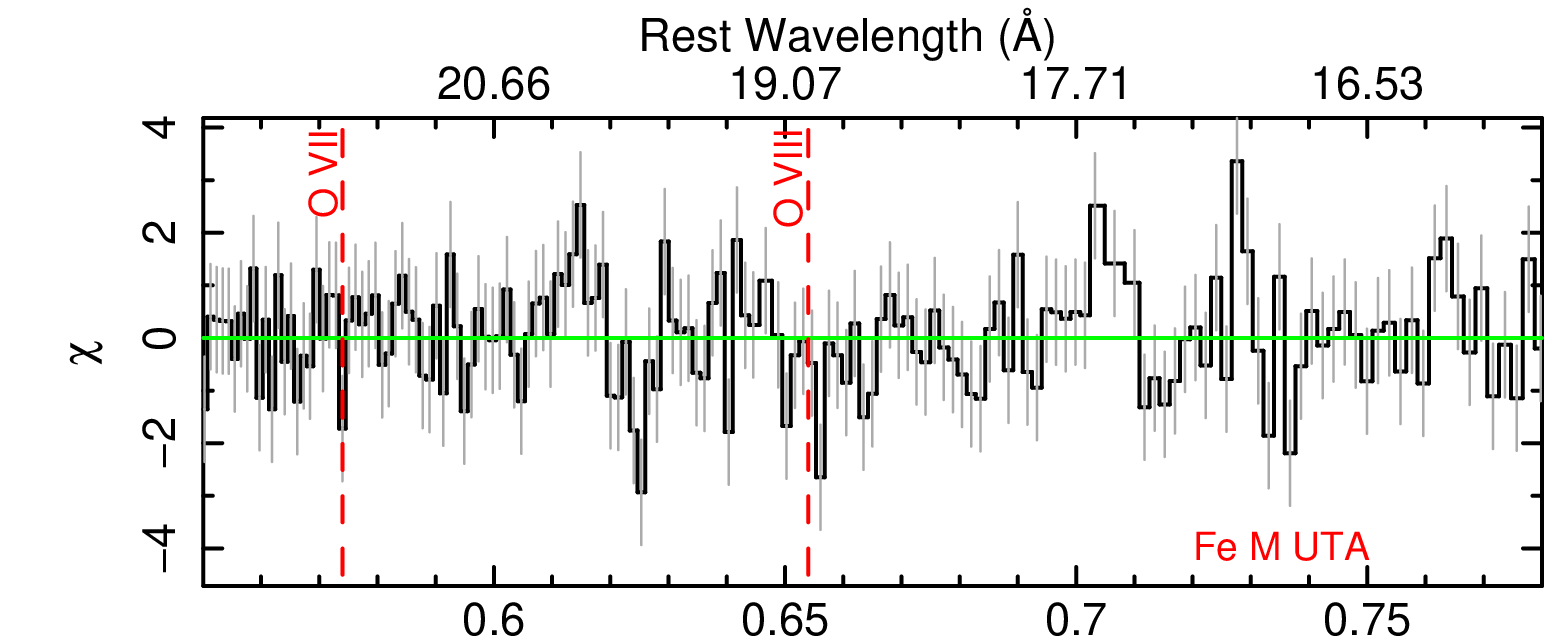}}
\rotatebox{-90}{\includegraphics[height=14cm]{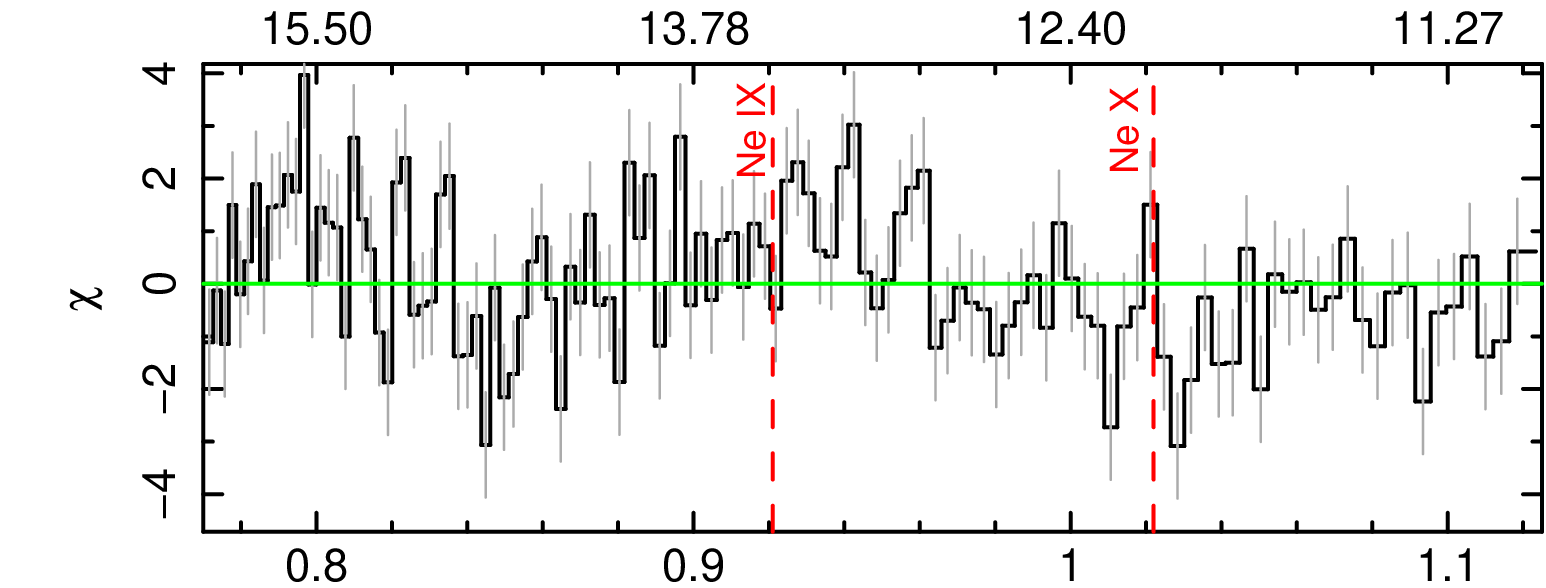}}
\rotatebox{-90}{\includegraphics[height=14cm]{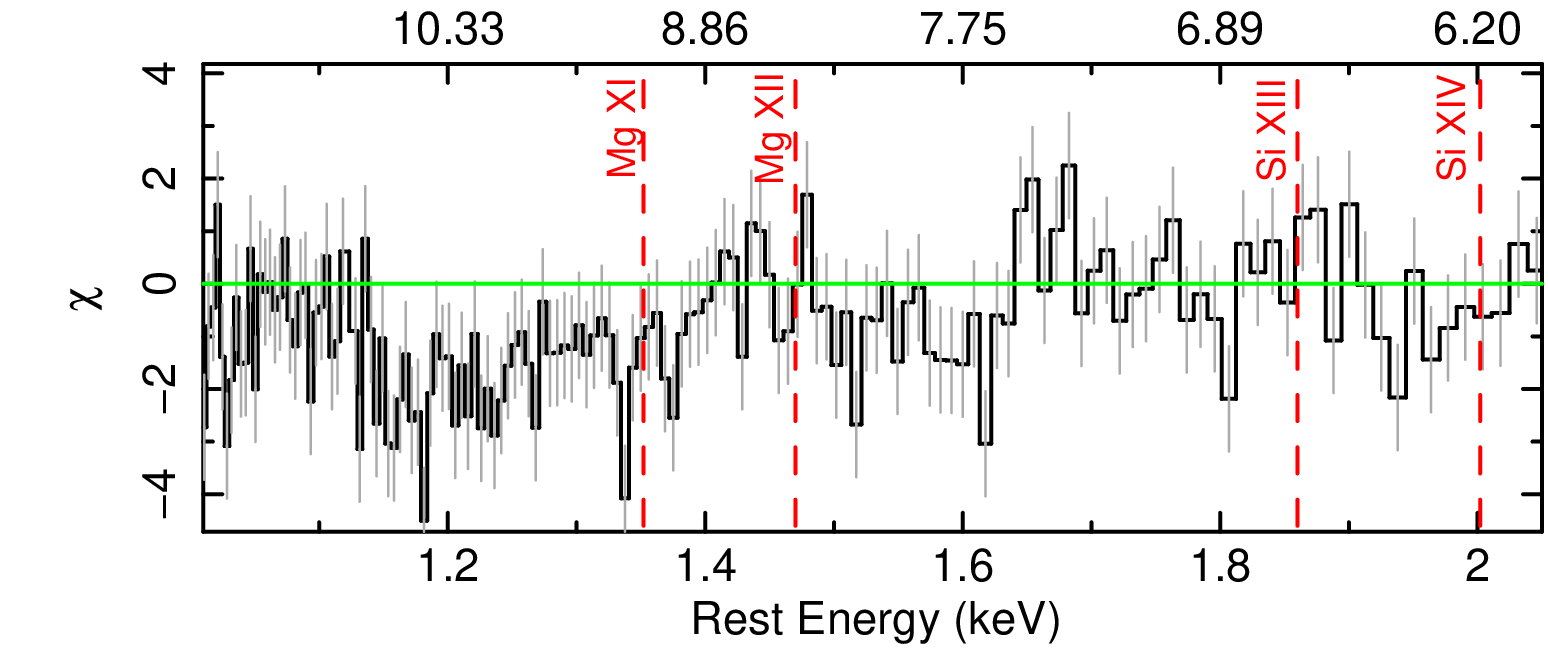}}
\end{center}
\caption{Deviations (in units $\sigma$) of the mean RGS data from the best-fit 
ISM absorption model obtained with the baseline \textsc{tbnew} model, as plotted 
in Figure 1 and listed in Table\,2. The $1\sigma$ errors are shown in greyscale.
The data have been rebinned with a finer resolution of $\Delta\lambda=0.05$\,\AA. 
Rest energy (in keV) is plotted along the lower x-axis, rest wavelength (in 
\AA) along the upper axis. Dashed lines mark the expected positions of the 
resonance components of the strongest He-like (He$\alpha$) and H-like (Ly$\alpha$) 
lines from abundant elements. The panels (from top to bottom) represent 
the residuals in the region the O, Ne and Mg/Si respectively. No significant 
narrow emission or absorption features are present near to the expected positions 
of the strongest atomic lines, indicating the lack of a low velocity warm absorber 
associated to PDS\,456, however clear broad residual structure is apparent, in 
particular in the 0.8--1.5\,keV range.}
\label{mean_panels}
\end{figure}

\clearpage

\begin{figure}
\begin{center}
\rotatebox{-90}{\includegraphics[height=14cm]{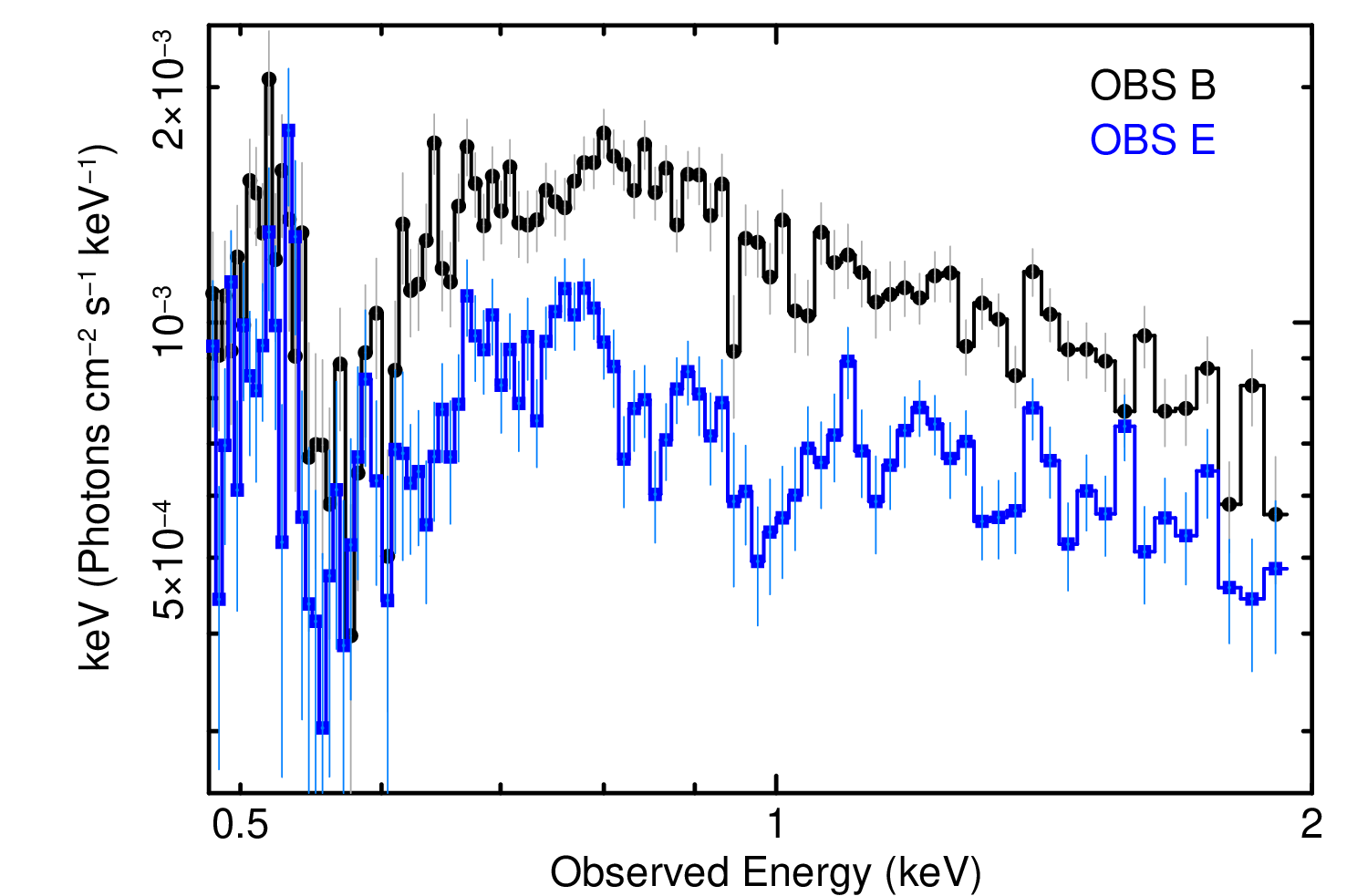}}
\rotatebox{-90}{\includegraphics[height=14cm]{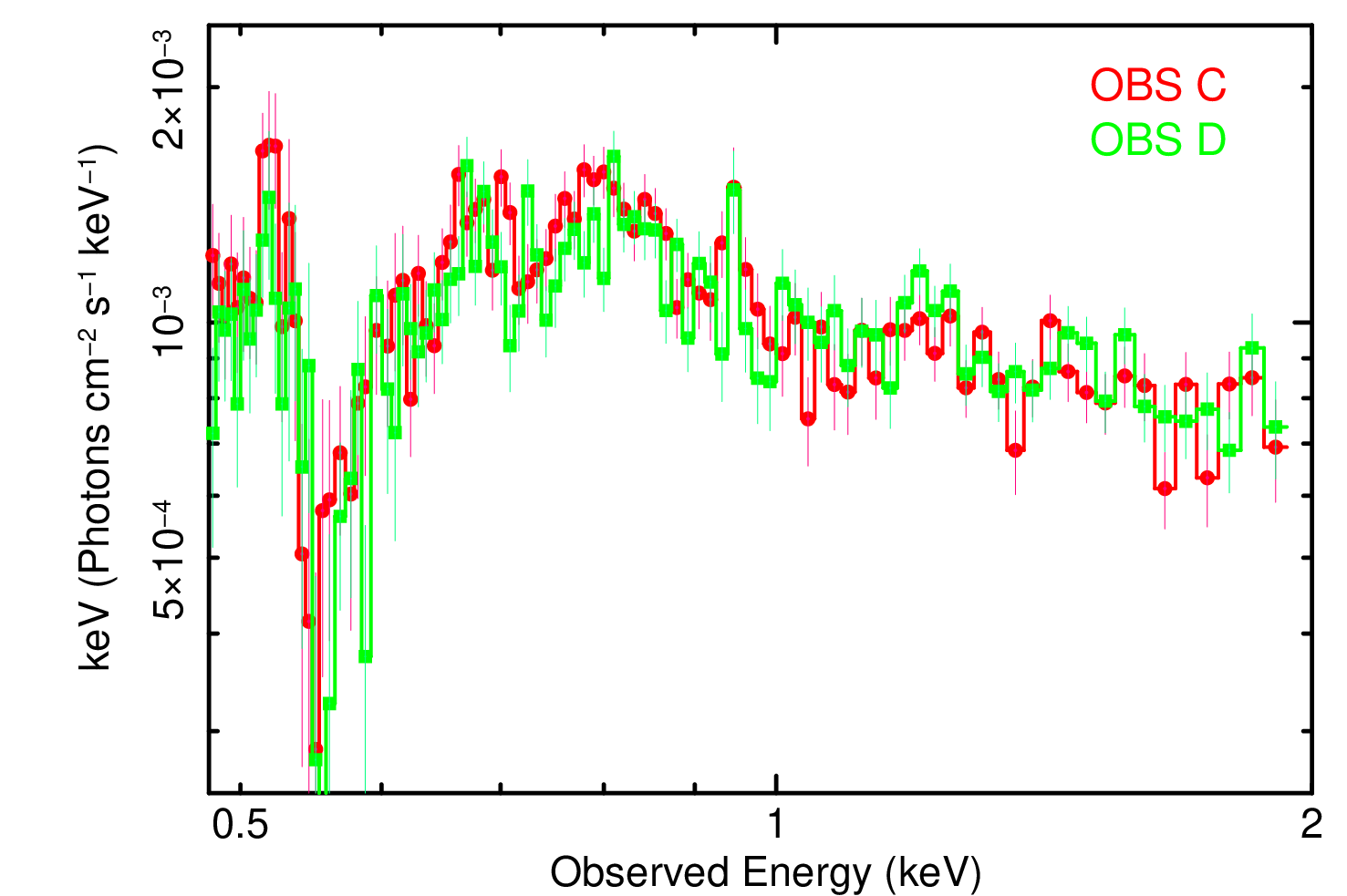}}
\end{center}
\caption{The four RGS spectral sequences of PDS\,456 taken during the 2013--14 \xmm\ 
campaign. The two panels show the fluxed spectra, unfolded against a simple powerlaw 
continuum. The upper plot shows the 2nd (OBS.\,B, black circles) and highest flux 
sequence versus the 5th (OBS.\,E, blue squares) and lowest flux sequences. Residual 
structure in the OBS.\,E spectrum is apparent around 1 keV in the observed frame. 
The lower panel shows the 3rd and 4th sequences (OBS.\,C and D, red circles and green squares respectively). Taken 
only 4 days apart, these are consistent in form, so they have been combined for the spectral 
analysis. The strong absorption between 0.5--0.6\,keV in all the spectra is associated 
to the neutral O\,\textsc{i} edge from our Galaxy.}
\label{fluxed}
\end{figure}

\clearpage

\begin{figure}
\begin{center}
\rotatebox{-90}{\includegraphics[height=12cm]{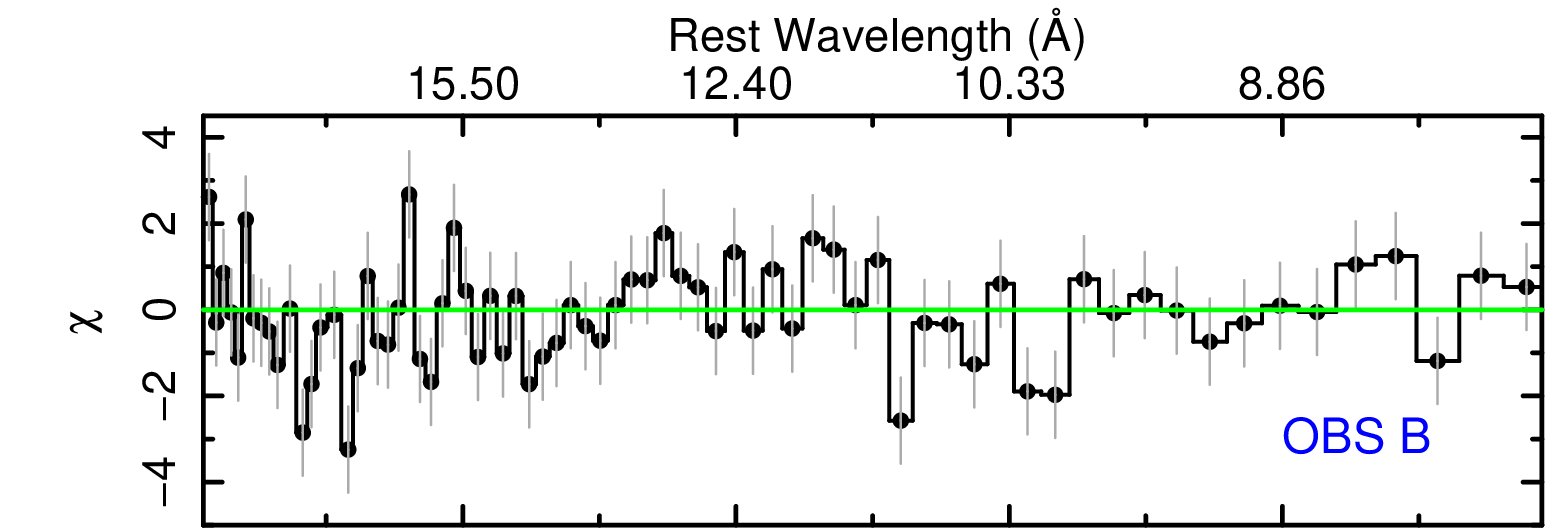}}
\rotatebox{-90}{\includegraphics[height=12cm]{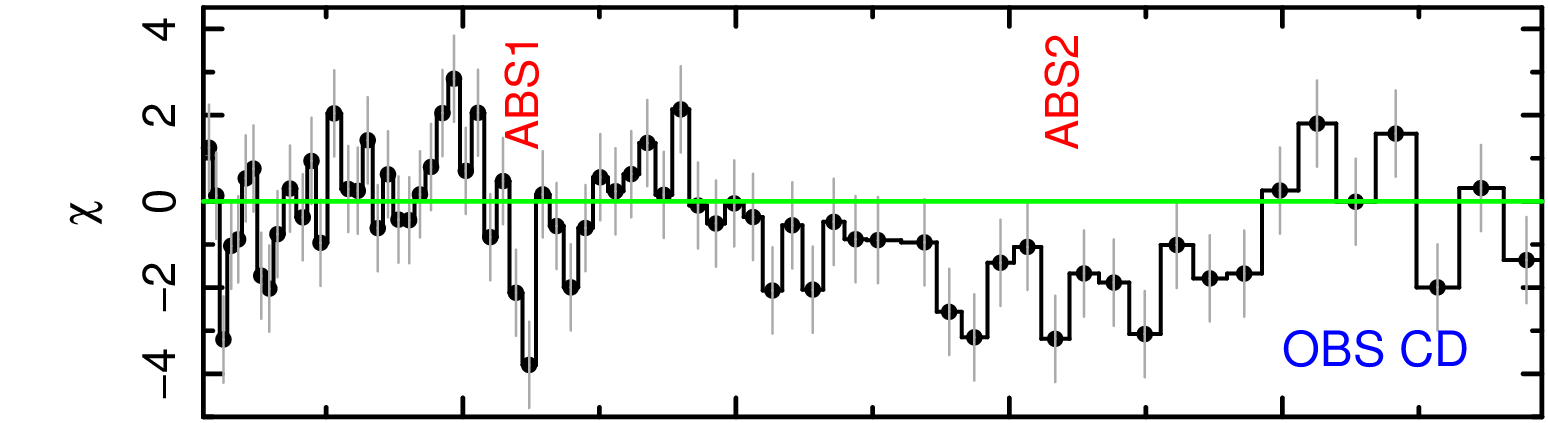}}
\rotatebox{-90}{\includegraphics[height=12cm]{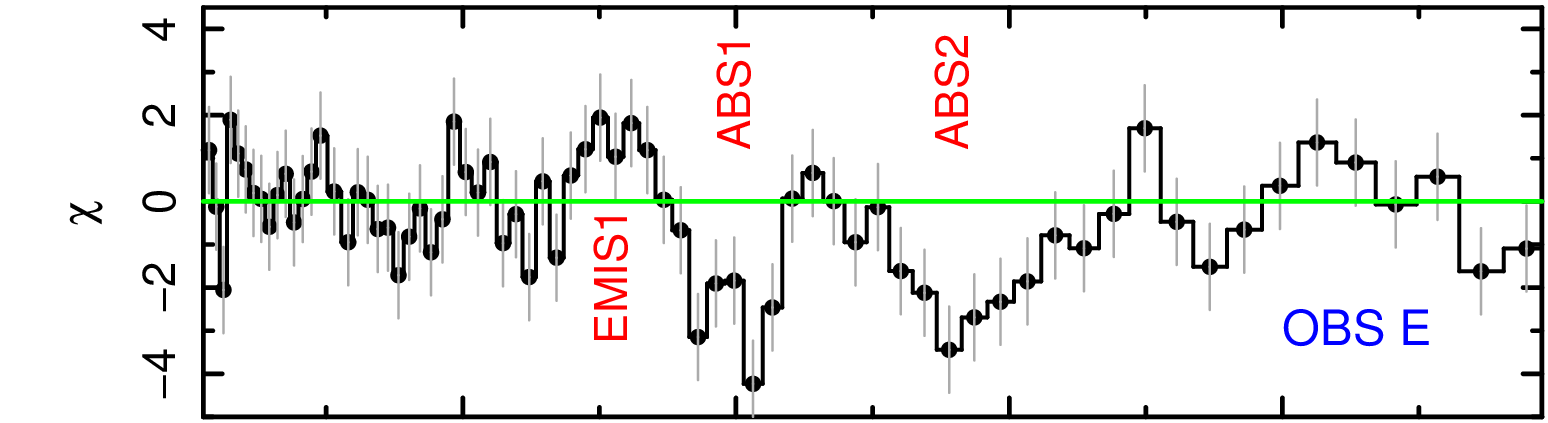}}
\rotatebox{-90}{\includegraphics[height=12cm]{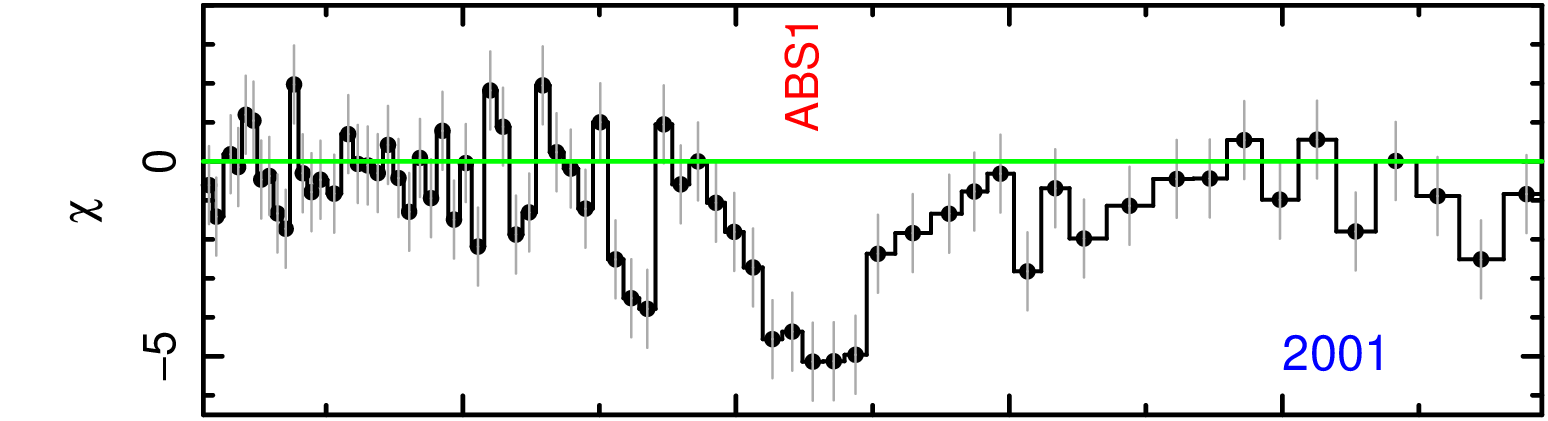}}
\rotatebox{-90}{\includegraphics[height=12cm]{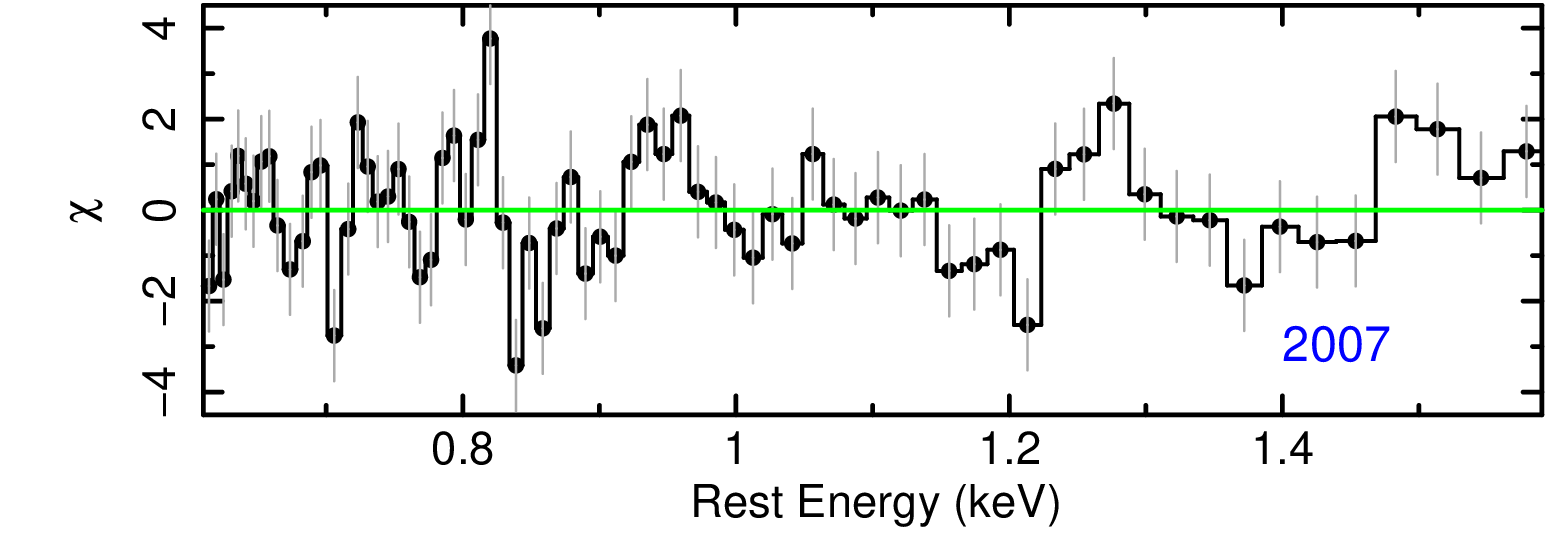}}
\end{center}
\caption{Panels showing the RGS spectral residuals to the baseline continuum model, 
for the 2013--2014 observations B, CD (combined), and E, as well as the earlier 2001 and 
2007 RGS observations of PDS\,456. The RGS spectra have been binned to 
$\Delta\lambda=0.2$\,\AA, at approximately twice the spectral resolution, and are 
plotted in the rest frame of PDS\,456 at $z=0.184$. Data points are shown as black 
circles, with $1\sigma$ error bars shown in grey. Broad residuals are particularly 
apparent in some of the spectral sequences, e.g. in the form of absorption troughs 
seen at 1\,keV and 1.17 keV in OBS.\,E, a deep absorption dip centered at 1.06\,keV in 
the 2001 sequence and a very broad absorption trough in OBS.\,CD near 1.2\,keV. 
The significant features labeled above are those fitted with Gaussian profiles in 
Table 2.}
\label{sequences}
\end{figure}

\clearpage

\begin{figure}
\begin{center}
\rotatebox{-90}{\includegraphics[height=14cm]{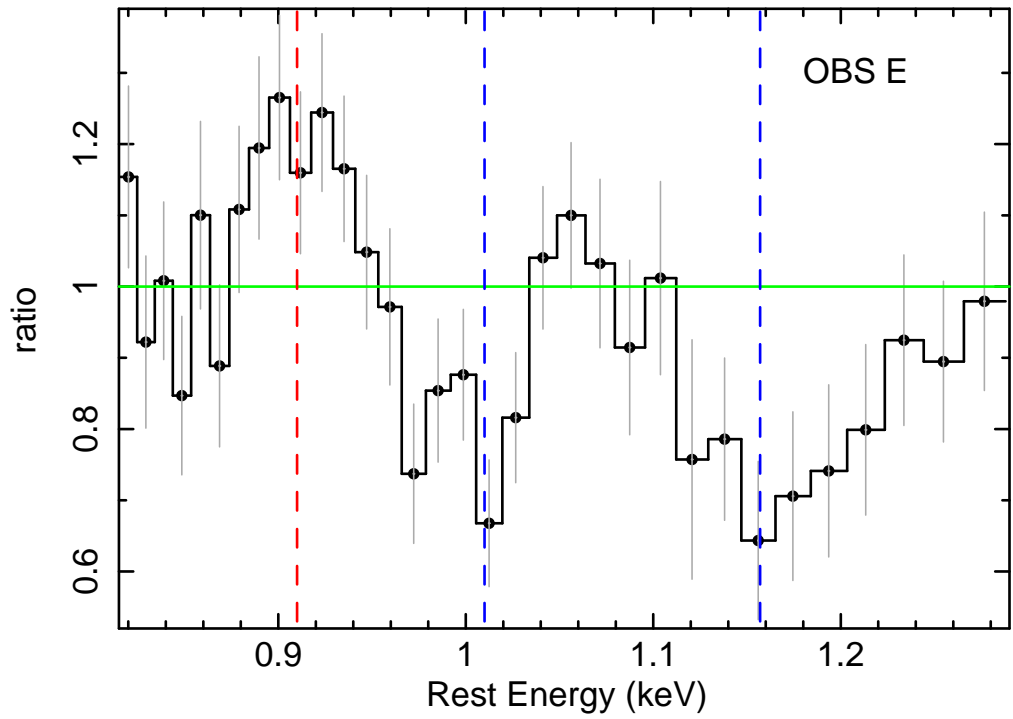}}
\end{center}
\caption{A zoom in of the RGS spectra for OBS.\,E, showing the complex soft X-ray 
features present near 1\,keV.  The spectrum is plotted as data/model residuals 
against the best-fit baseline continuum versus rest frame energy. The approximate 
energy centroids are marked as a red dashed line for the significant emission line 
and in blue for the significant absorption lines (see Table 2).}
\label{pcyg}
\end{figure}

\clearpage

\begin{figure}
\begin{center}
\rotatebox{-90}{\includegraphics[height=12cm]{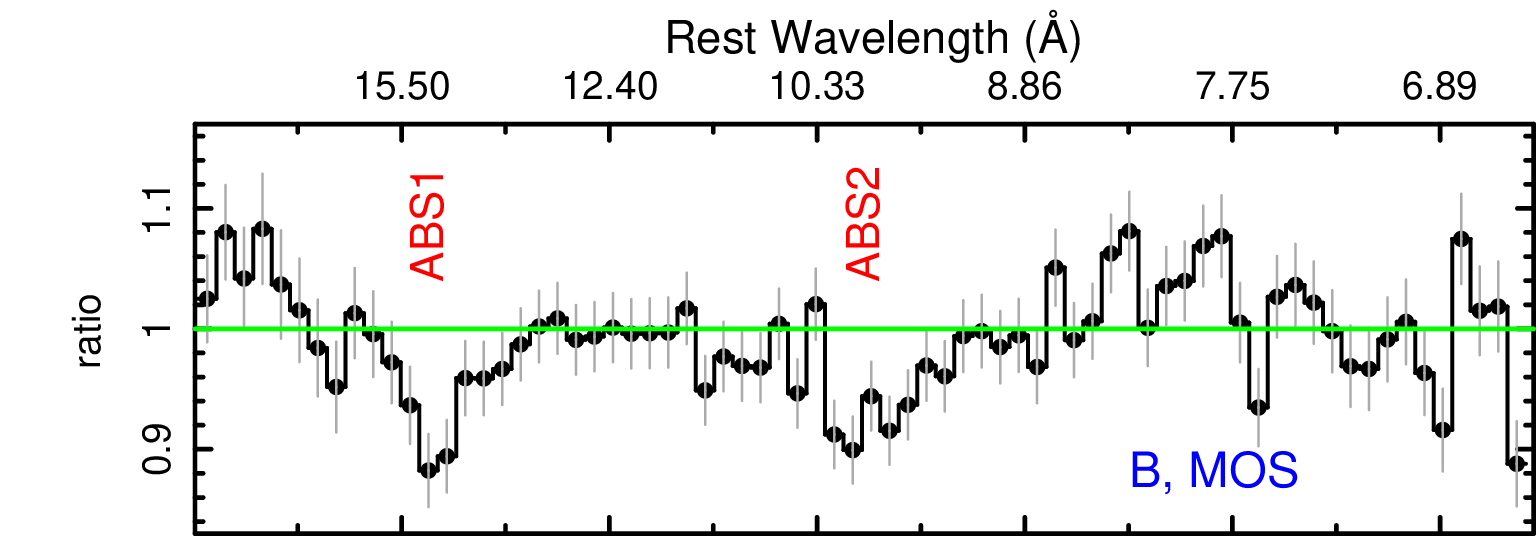}}
\rotatebox{-90}{\includegraphics[height=12cm]{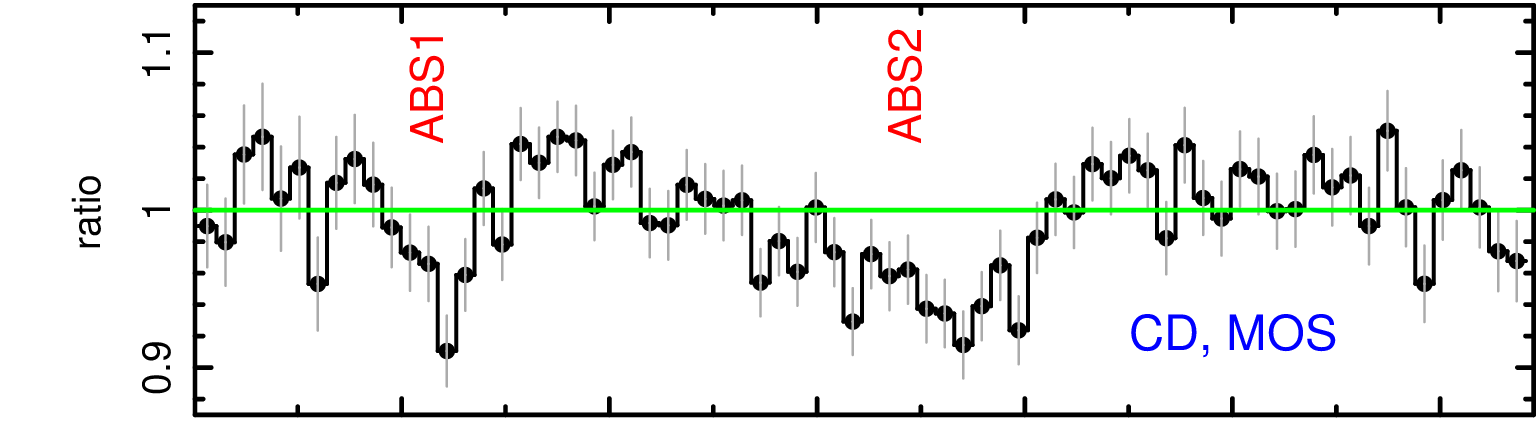}}
\rotatebox{-90}{\includegraphics[height=12cm]{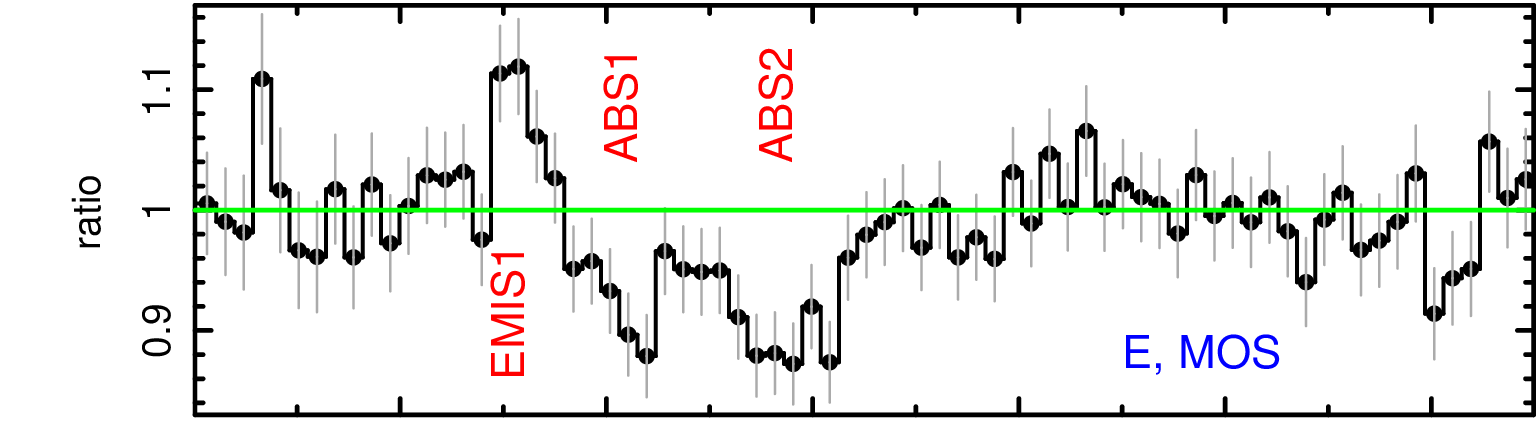}}
\rotatebox{-90}{\includegraphics[height=12cm]{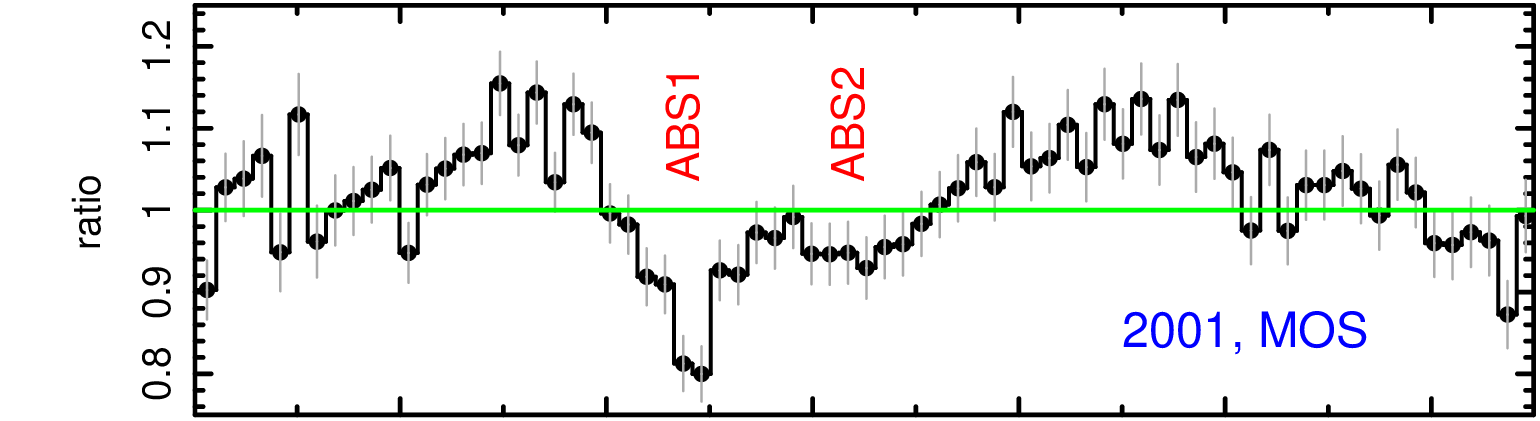}}
\rotatebox{-90}{\includegraphics[height=12cm]{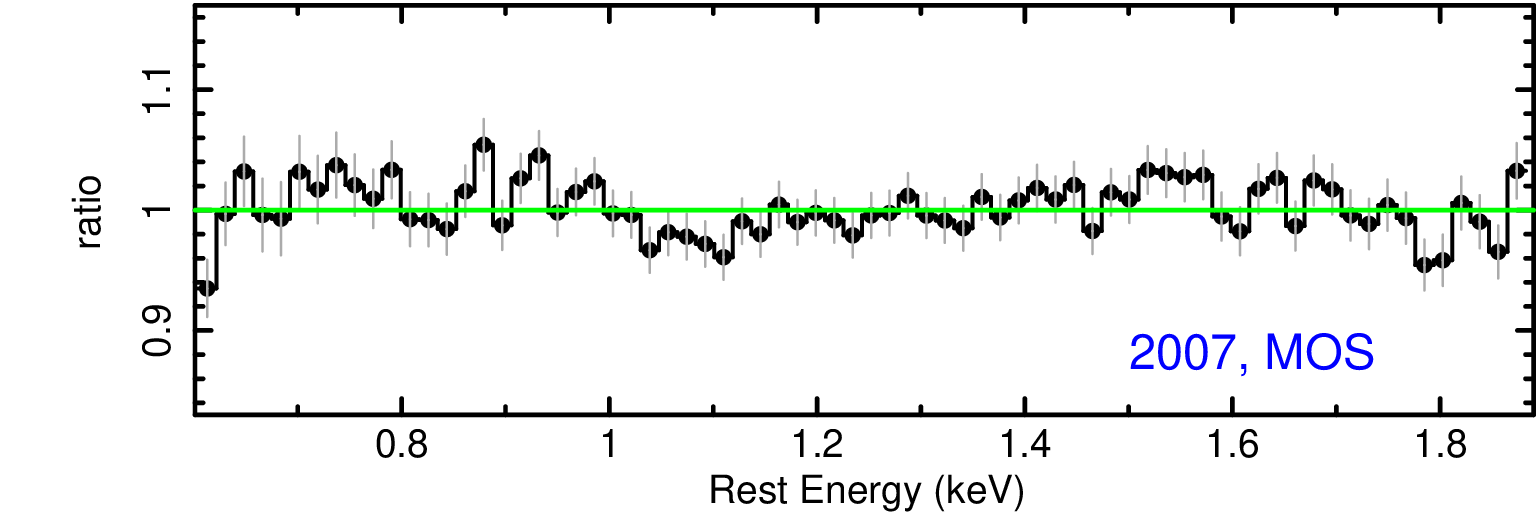}}
\end{center}
\caption{Data/model residuals in the soft X-ray band to the EPIC MOS 
observations, showing from top to bottom the 2013--2014 observations B, CD, and E, 
as well as the earlier spectra from 2001 and 2007. The baseline continuum is a 
power-law plus blackbody component, absorbed by a Galactic column, as per the RGS. 
Several line features that are present in the RGS are also confirmed in the MOS 
spectra; in particular, note the two absorption troughs between 1.0--1.2\,keV in OBS.\,E, 
the deep absorption profile above 1\,keV in the 2001 observation, and the broad absorption 
trough above 1.2\,keV in OBS.\,CD. The significant features labeled above are those fitted 
with Gaussian profiles in Table 3.
}
\label{MOS}
\end{figure}

\clearpage
\begin{figure}
\begin{center}
\rotatebox{-90}{\includegraphics[height=12cm]{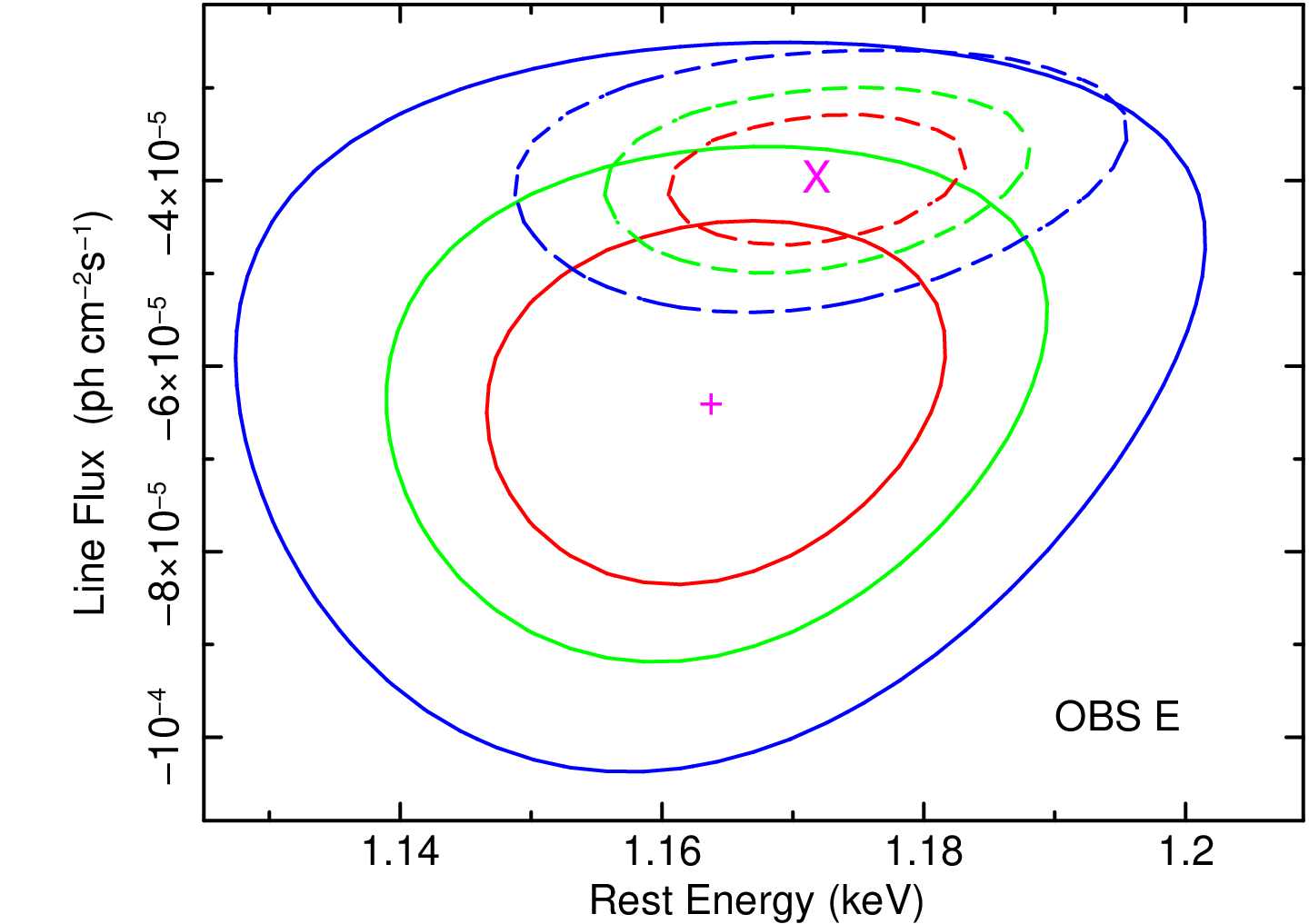}}
\rotatebox{-90}{\includegraphics[height=12cm]{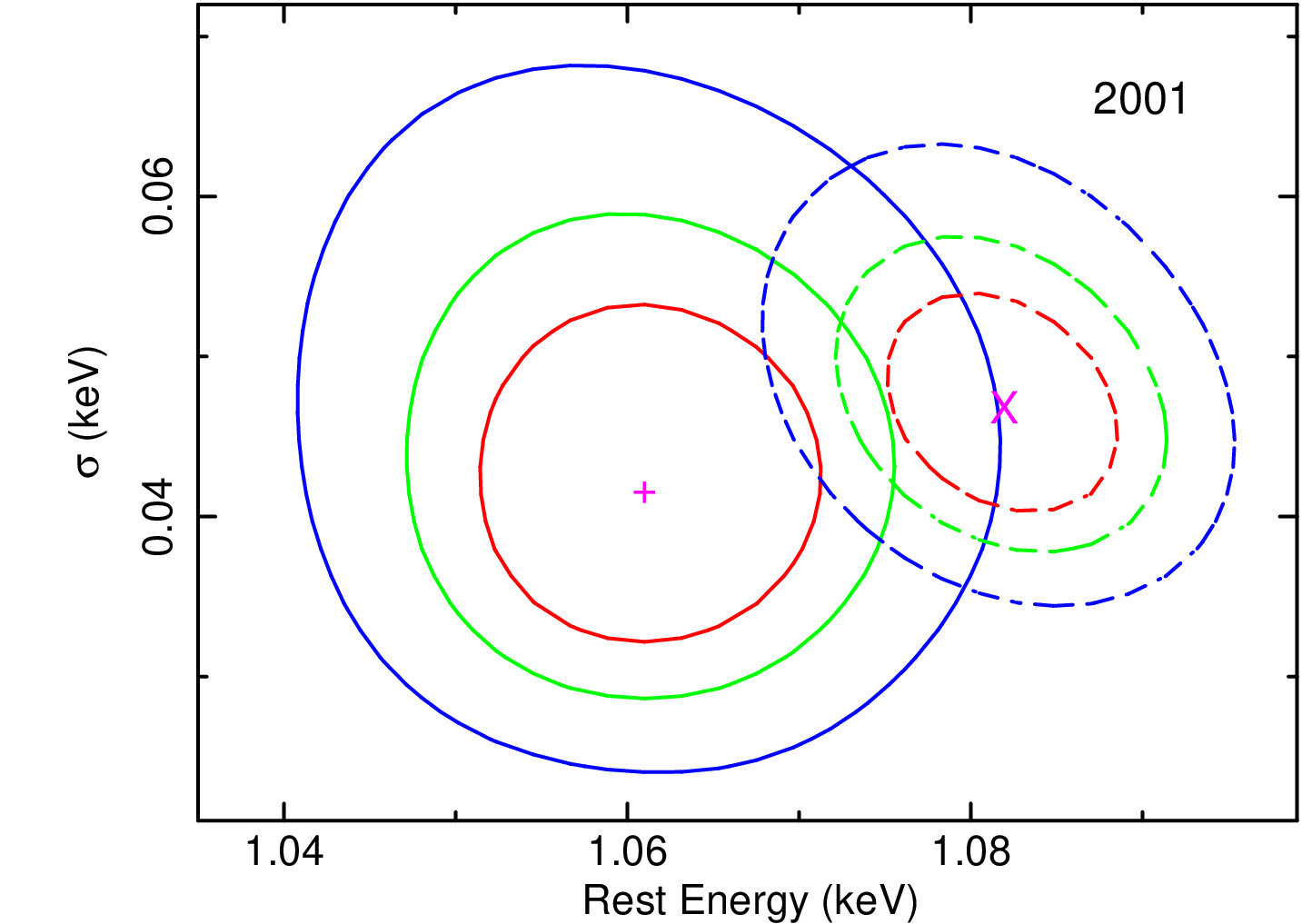}}
\end{center}
\caption{Contour plots, showing the comparison between two of the broad absorption lines detected in 
RGS and MOS spectra. The contours for the RGS are shown as solid lines, while 
the MOS is indicated by dot-dashed lines. The contours represent the 68\%, 90\% and 99\% levels 
for two parameters of interest. The upper panel shows the contours between rest energy and line flux 
for the 1.16\,keV absorption line in the OBS.\,E 
spectrum, illustrating the independent agreement between the two spectra. The lower panel shows the 
contours for the broad absorption trough in the 2001 spectrum, showing the agreement in the 
width of the line profile ($\sigma$) measured between the two detectors.}
\label{contours}
\end{figure}

\clearpage

\begin{figure}
\begin{center}
\rotatebox{-90}{\includegraphics[height=12cm]{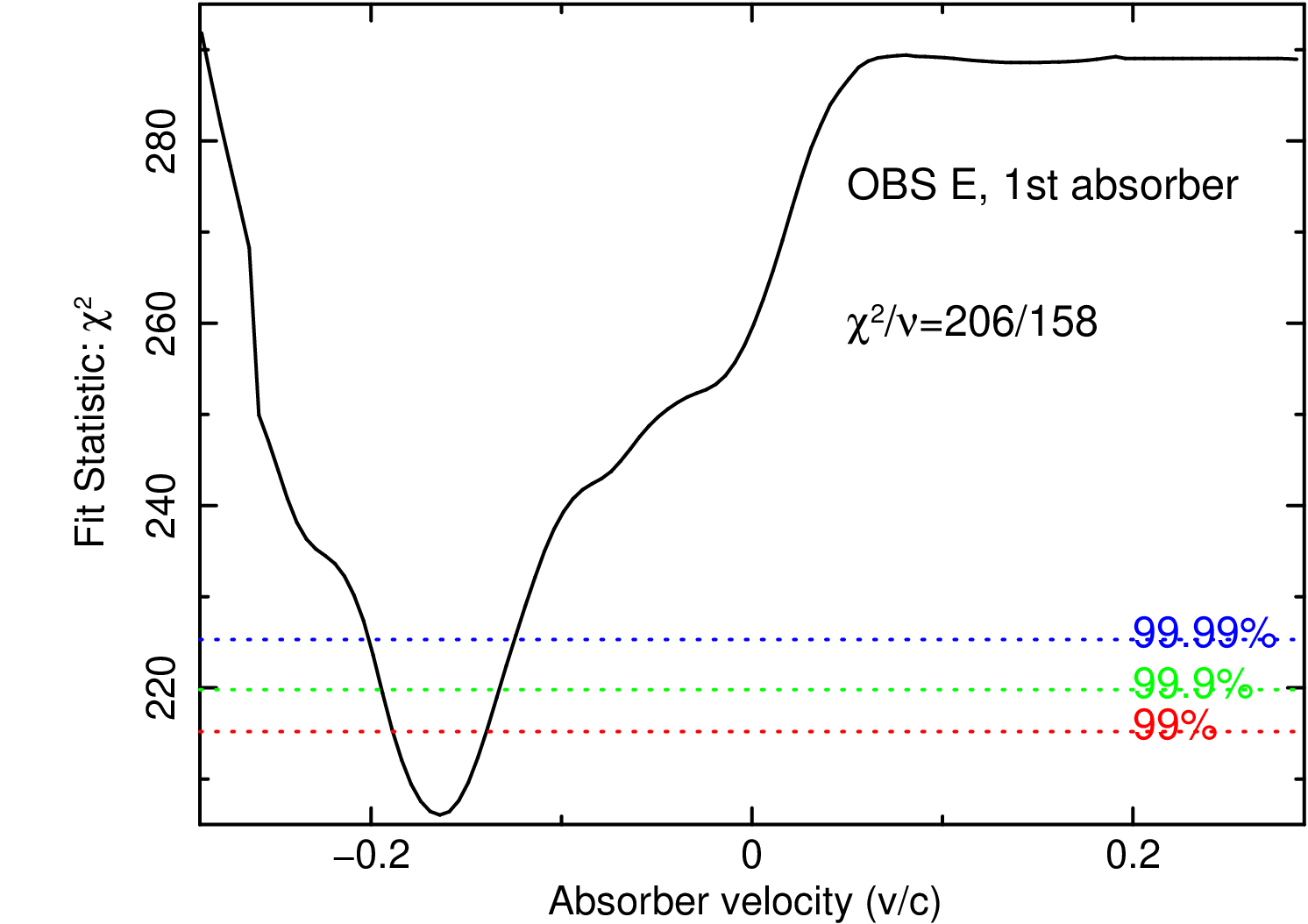}}
\rotatebox{-90}{\includegraphics[height=12cm]{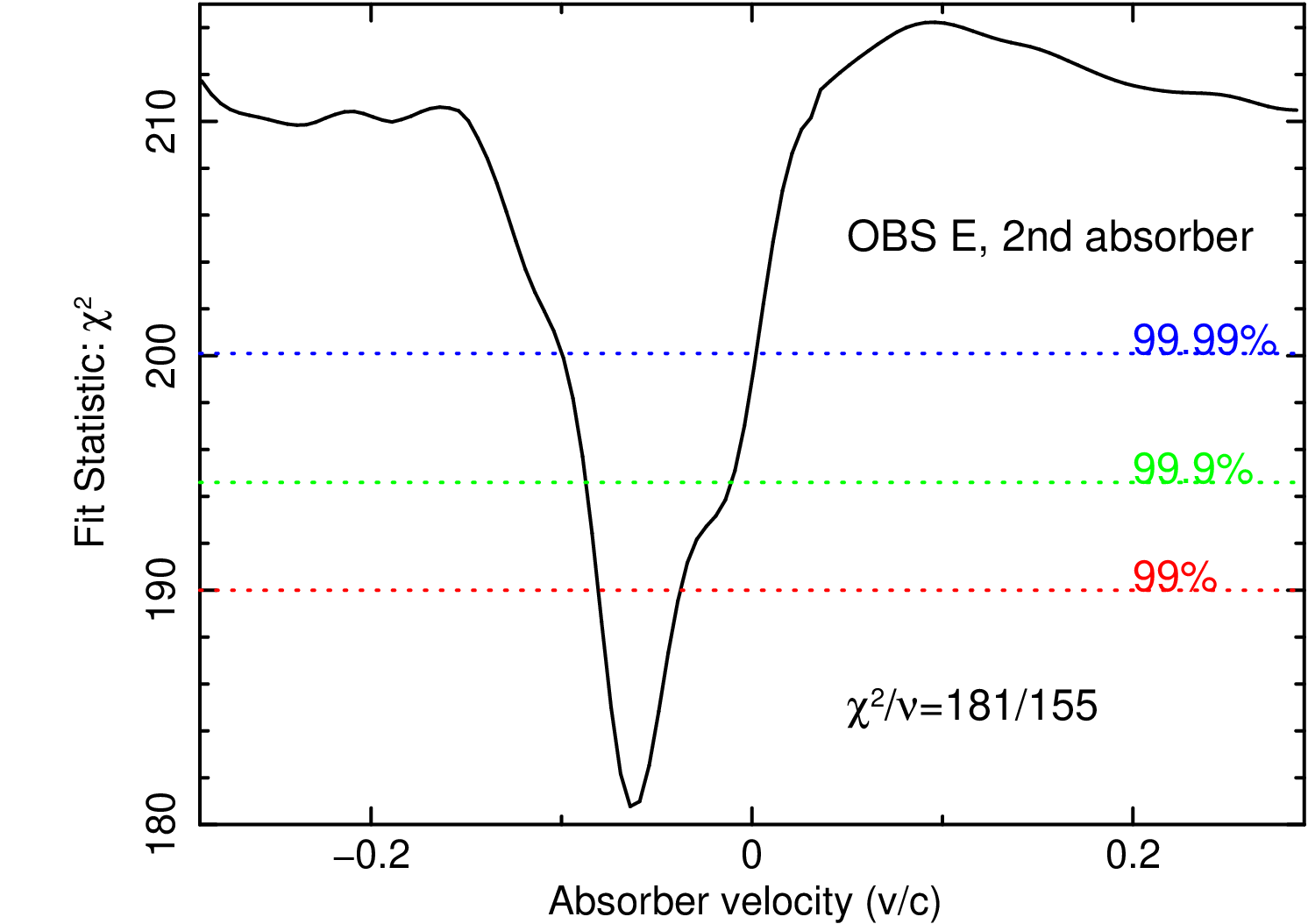}}
\end{center}
\caption{Fit minimization in terms of $\chi^{2}$ vs velocity for the OBS.\,E RGS+MOS spectra. 
Here a negative velocity represents outflow or blue-shift.
The plots show the changes in the resultant $\chi^{2}$ after stepping through 
velocity space for the addition of two successive absorption zones to the model, as described in the text.
The upper panel shows a well determined minimum at an outflow velocity of $v_{\rm out}=-0.17c$ for 
the fast absorption zone. After accounting for the fast zone in the fit, 
the lower panel shows a further minimum of $v_{\rm out}=-0.06c$ 
corresponding to an additional slower absorption zone. The horizontal dotted lines correspond 
to the 99\%, 99.9\% and 99.99\% confidence intervals for 2 parameters of interest. 
Note a solution with zero velocity or redshift is clearly excluded by the data. The resultant 
best-fit reduced chi-squared upon the addition of each absorption zone is also shown in each panel.}
\label{velocity}
\end{figure}

\clearpage

\begin{figure}
\begin{center}
\rotatebox{-90}{\includegraphics[height=14cm]{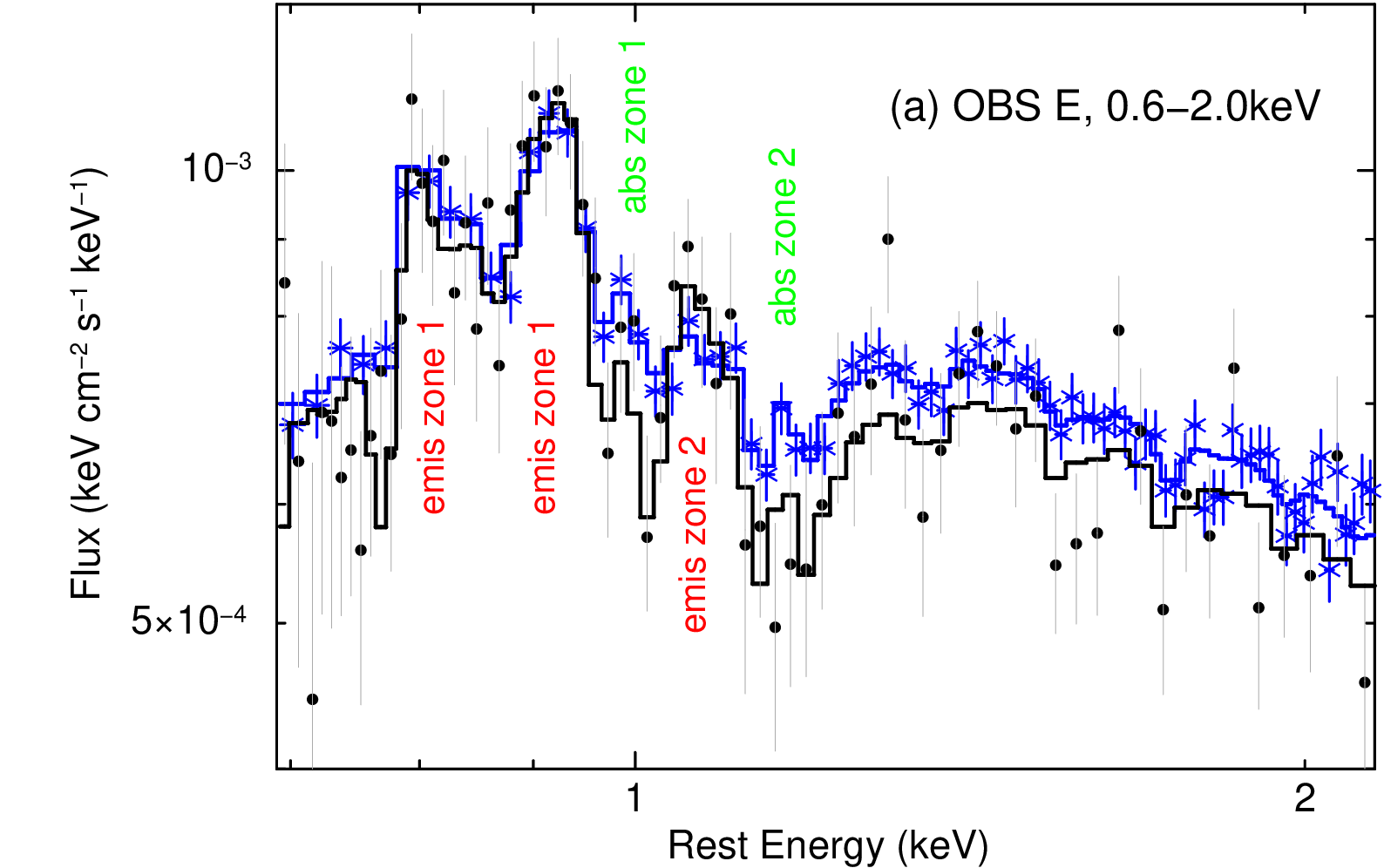}}
\rotatebox{-90}{\includegraphics[height=14cm]{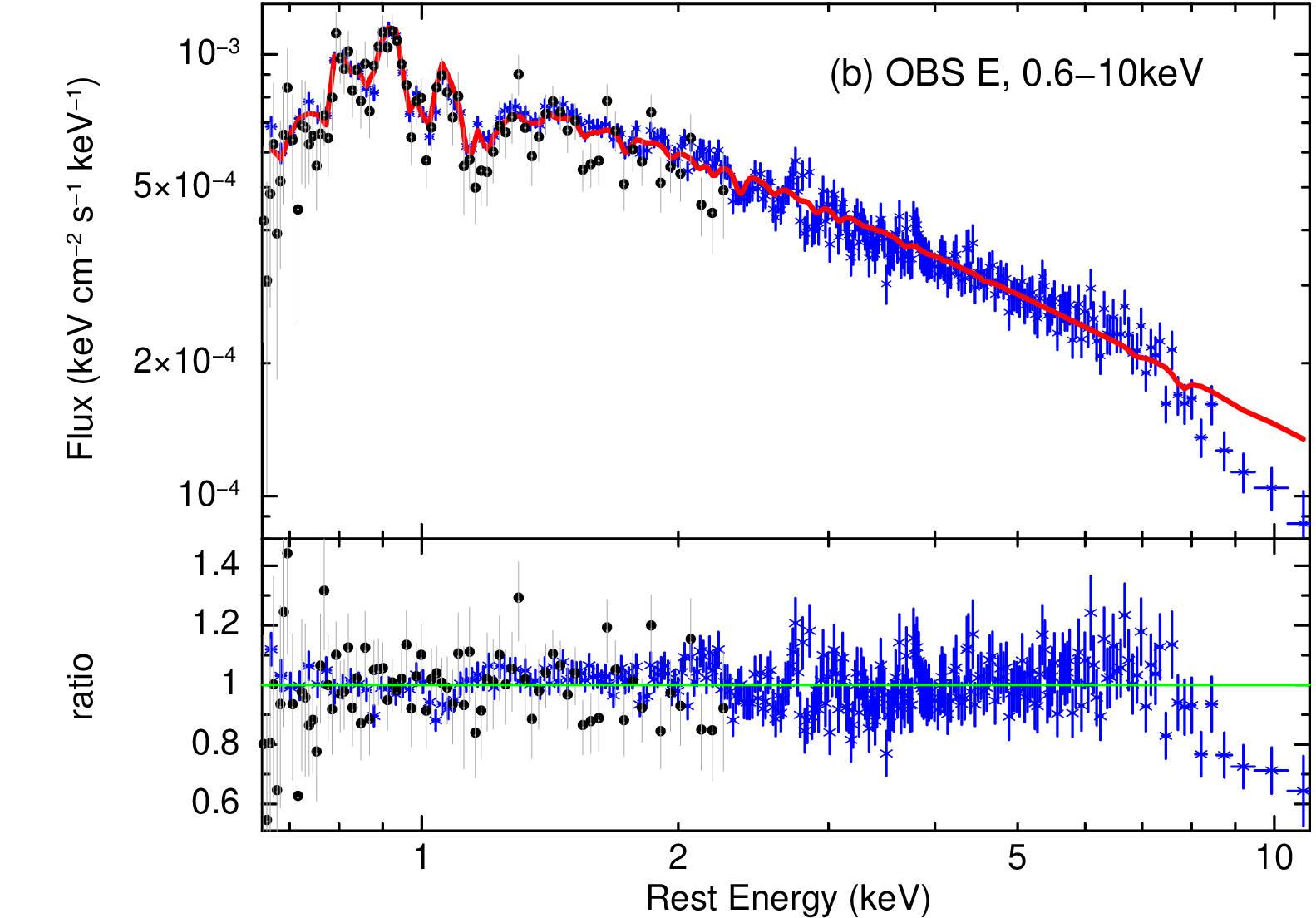}}
\end{center}
\caption{Best fit {\sc xstar} absorption model applied to observation E. The top 
panel shows the {\sc xstar} model fitted simultaneously from 0.6--2.0 keV to the MOS 
(data points, blue stars) and RGS (black circles), while the best-fit model fitted 
to each instrument is a solid line. The \textsc{xstar} model is able to account for 
the absorption structure with 2 different velocity zones of absorbing gas (absorption 
zones 1 and 2 in Table\,4) and the excess emission with 2 corresponding zones (see 
Table 5). 
The lower panel shows the same model, but extrapolated up to 10 keV in the 
MOS. The soft X-ray absorption model is able to account for the broad-band 
spectrum up to 10 keV, with the exception of the P-Cygni like absorption feature 
at iron K.}
\label{xstar}
\end{figure}

\clearpage
\begin{figure}
\begin{center}
\rotatebox{-90}{\includegraphics[height=14cm]{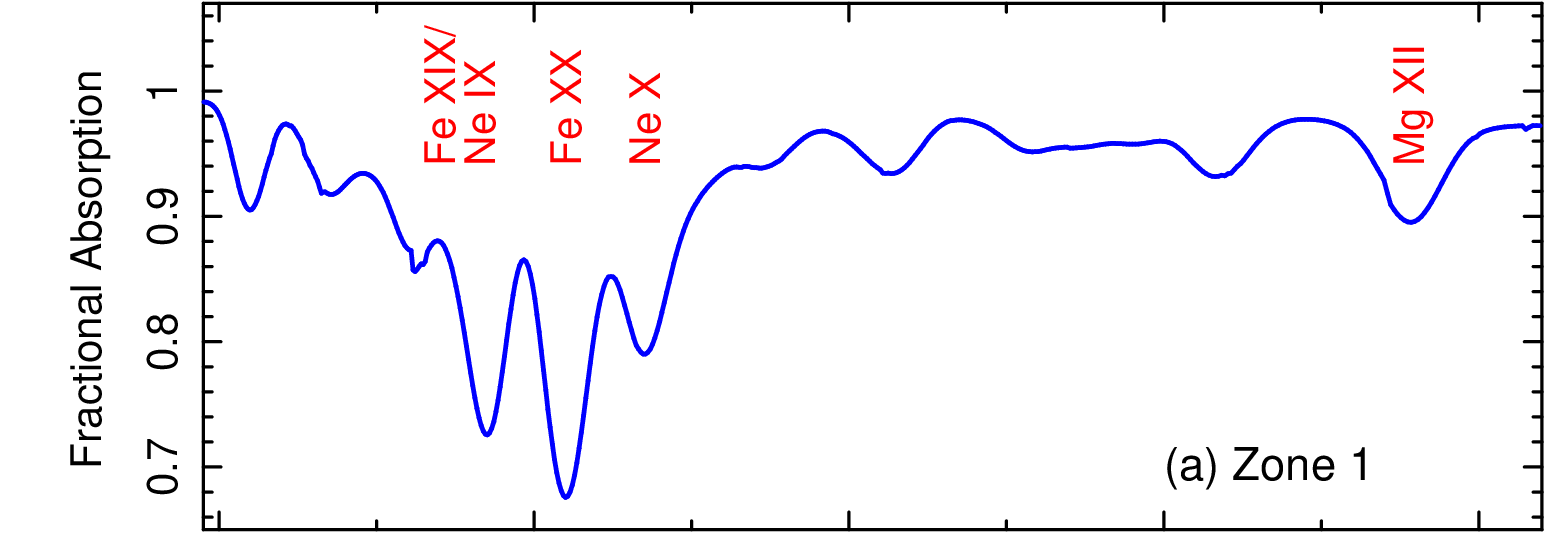}}
\rotatebox{-90}{\includegraphics[height=14cm]{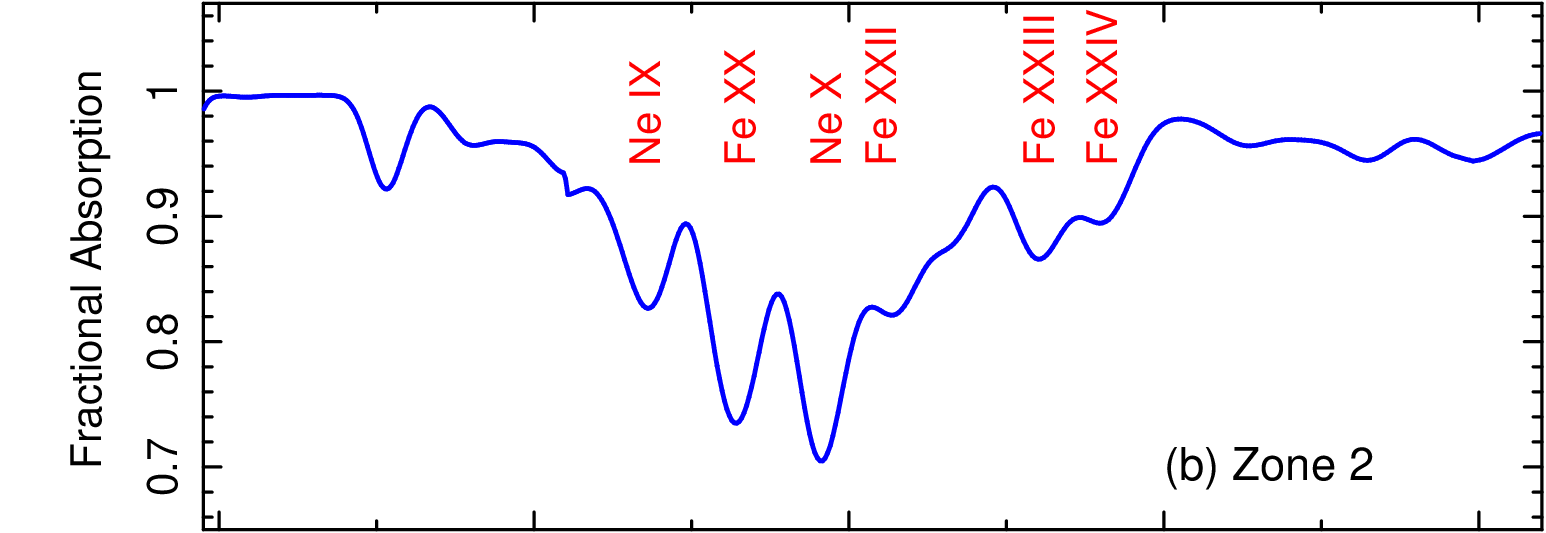}}
\rotatebox{-90}{\includegraphics[height=14cm]{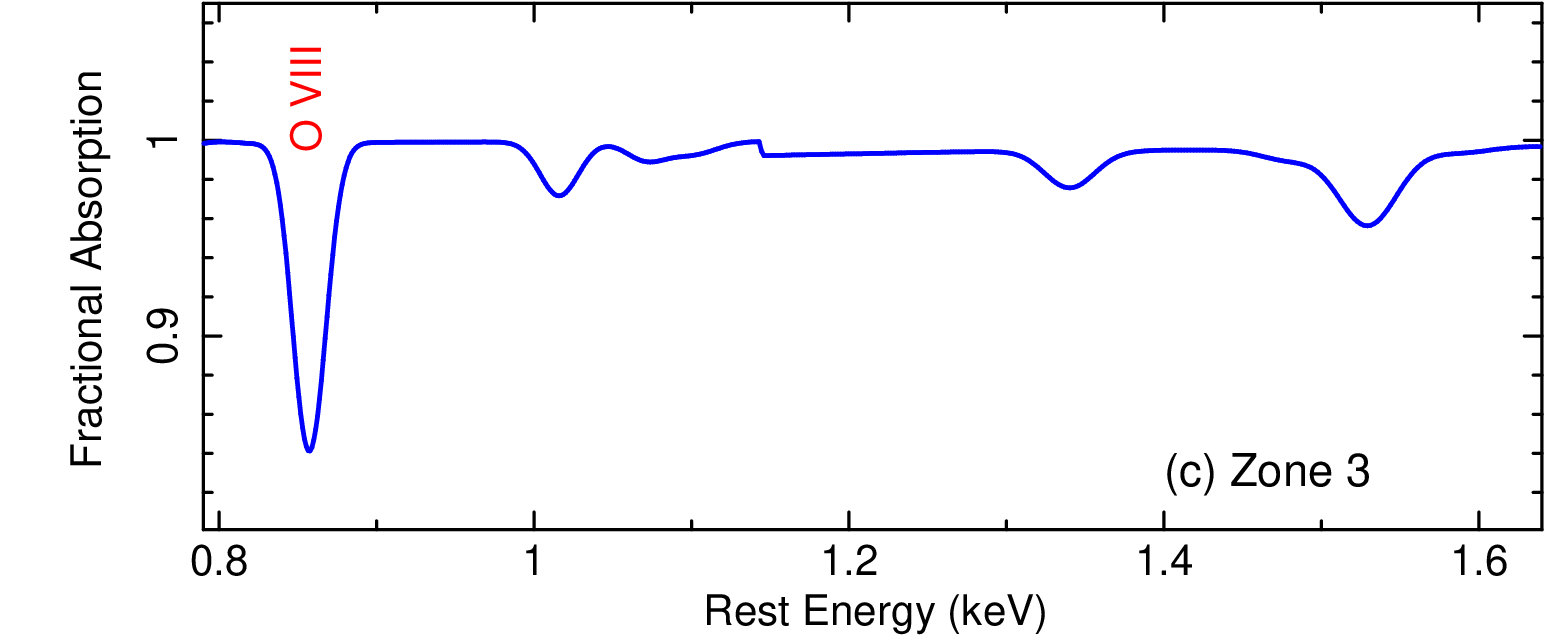}}
\end{center}
\caption{A graphical representation of the three {\sc xstar} absorber zones (see 
Table 4), showing the fraction of the continuum absorbed by each zone in the region 
around 1\,keV. Zone 1 (upper panel), corresponds to the lowest velocity ($0.06\,c$), 
lower ionization component in OBS.\,E, which mainly produces the 1 keV absorption 
trough seen in this observation. Zone 2 (middle panel) is a faster zone, and the 
blend of lines is responsible for the broad absorption trough observed near 1.2 keV. 
Most of the contribution towards zones 1 and 2 arises from the L-shell lines of iron 
and K-shell Ne. Zone 3 (lower panel) is the highest ionization and fastest zone 
present in OBS.\,CD, which makes only a small contribution towards the soft band, 
as it is mainly responsible for the blueshifted absorption near to 0.8\,keV, which 
is due to O\,\textsc{viii} Ly$\alpha$.}
\label{zones}
\end{figure}

\clearpage
\begin{figure}
\begin{center}
\rotatebox{-90}{\includegraphics[height=14cm]{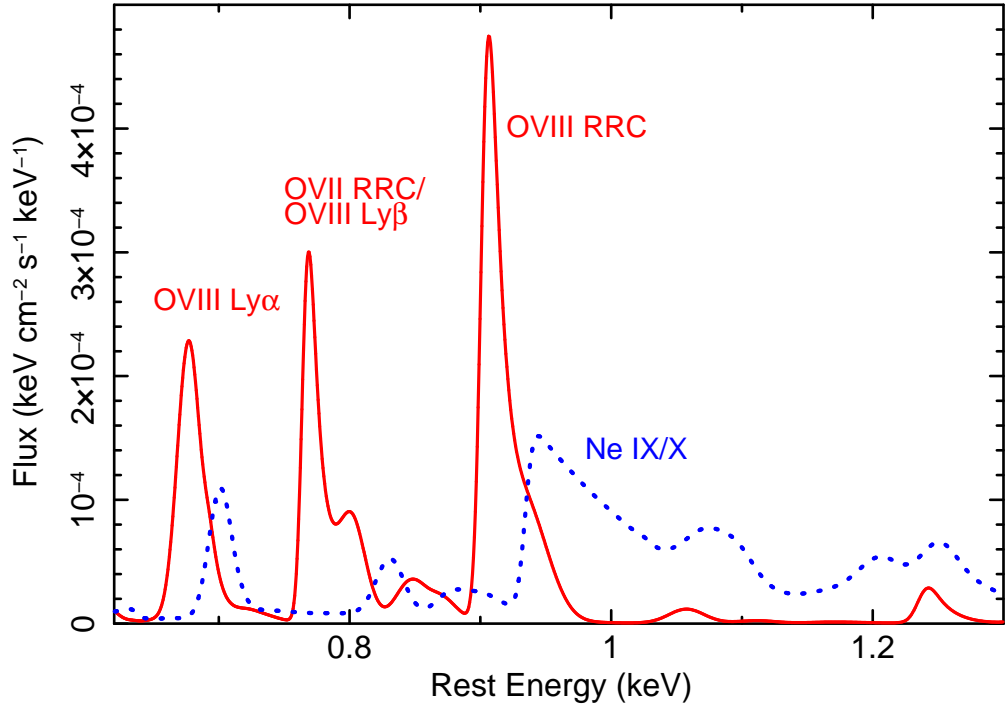}}
\end{center}
\caption{The relative contribution of the two soft X-ray photoionized emission zones 
from {\it XMM--Newton} OBS.\,E, as modeled by \textsc{xstar} with a velocity broadening 
of $\sigma=5,000$\,km\,s$^{-1}$ (see Table 5). The solid (red) line represents the lower 
ionization ($\log\xi=2.8$) emission zone 1, which contributes towards the majority of the 
soft X-ray line emission. In particular, the strongest line observed at $\sim0.9$\,keV 
likely originates from the blueshifted O\,\textsc{viii} RRC, while at lower energies the 
O\,\textsc{viii} Ly$\alpha$ line is comparatively weaker, as it is suppressed by Galactic 
absorption. The dashed (blue) line illustrates the much weaker contribution that the 
higher ionization zone 2 makes towards the emission spectrum.}
\label{emission}
\end{figure}

\clearpage

\begin{figure}
\begin{center}
\includegraphics[angle=-90,width=0.49\textwidth]{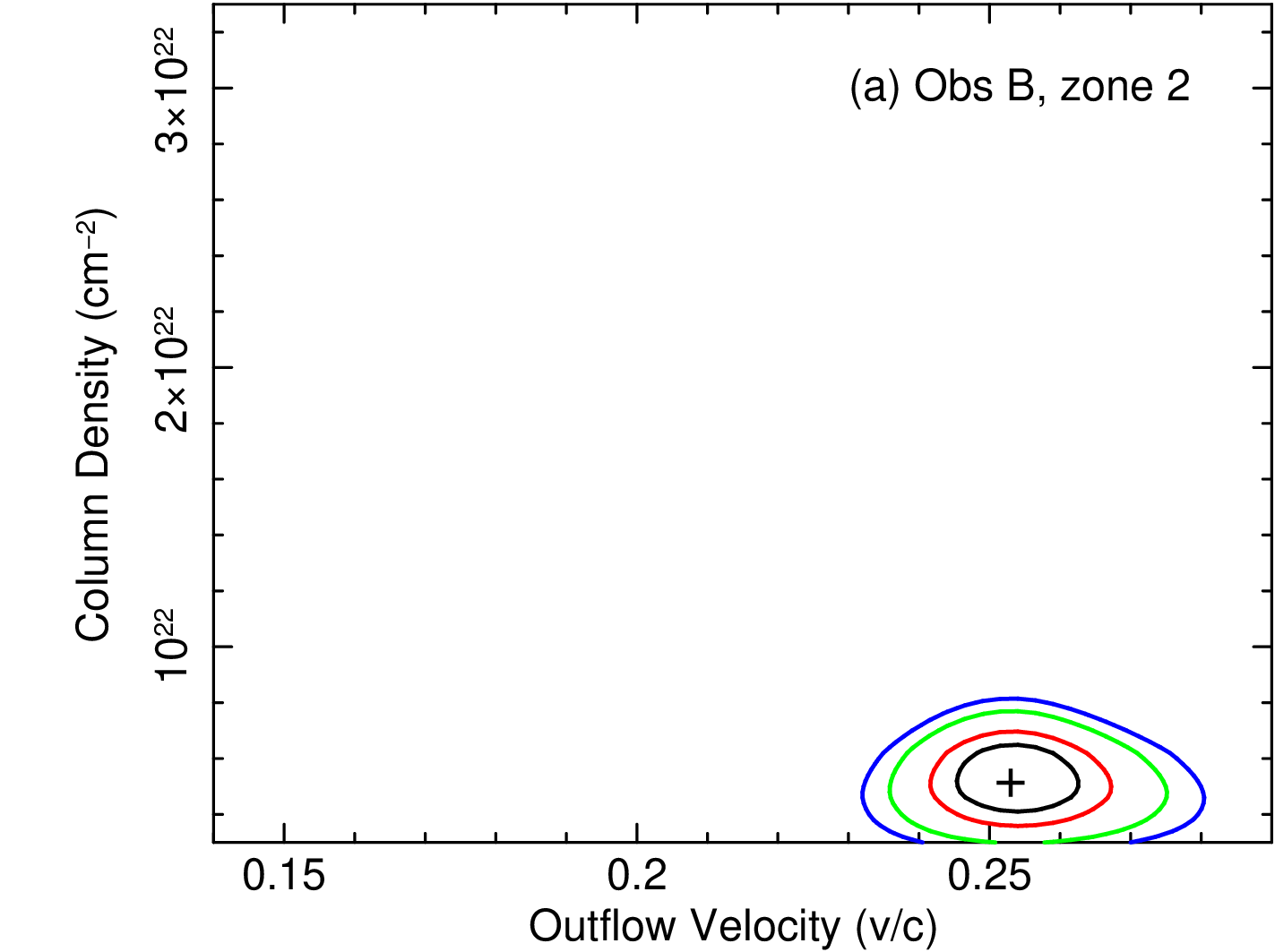}
\includegraphics[angle=-90,width=0.49\textwidth]{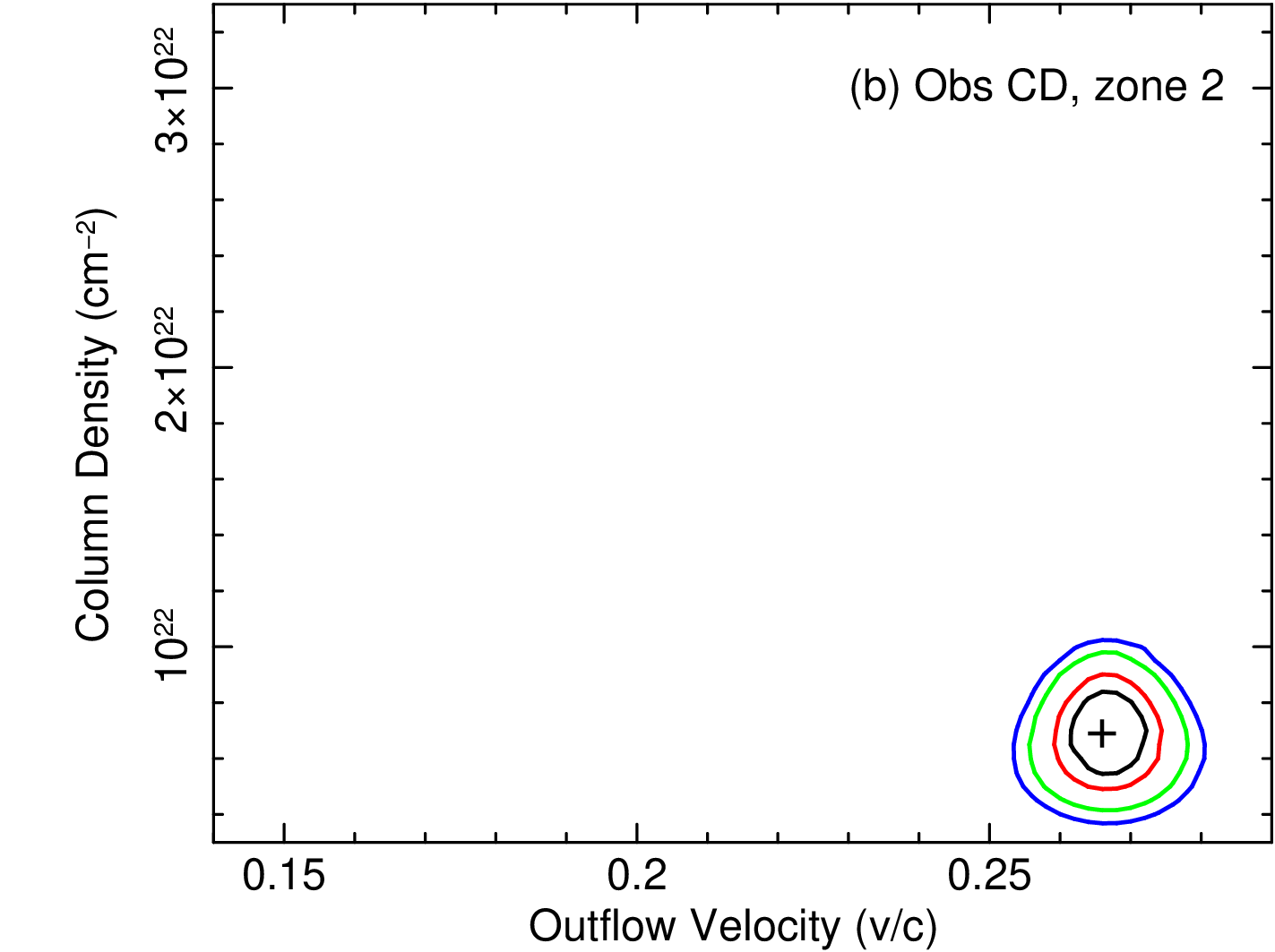}
\includegraphics[angle=-90,width=0.49\textwidth]{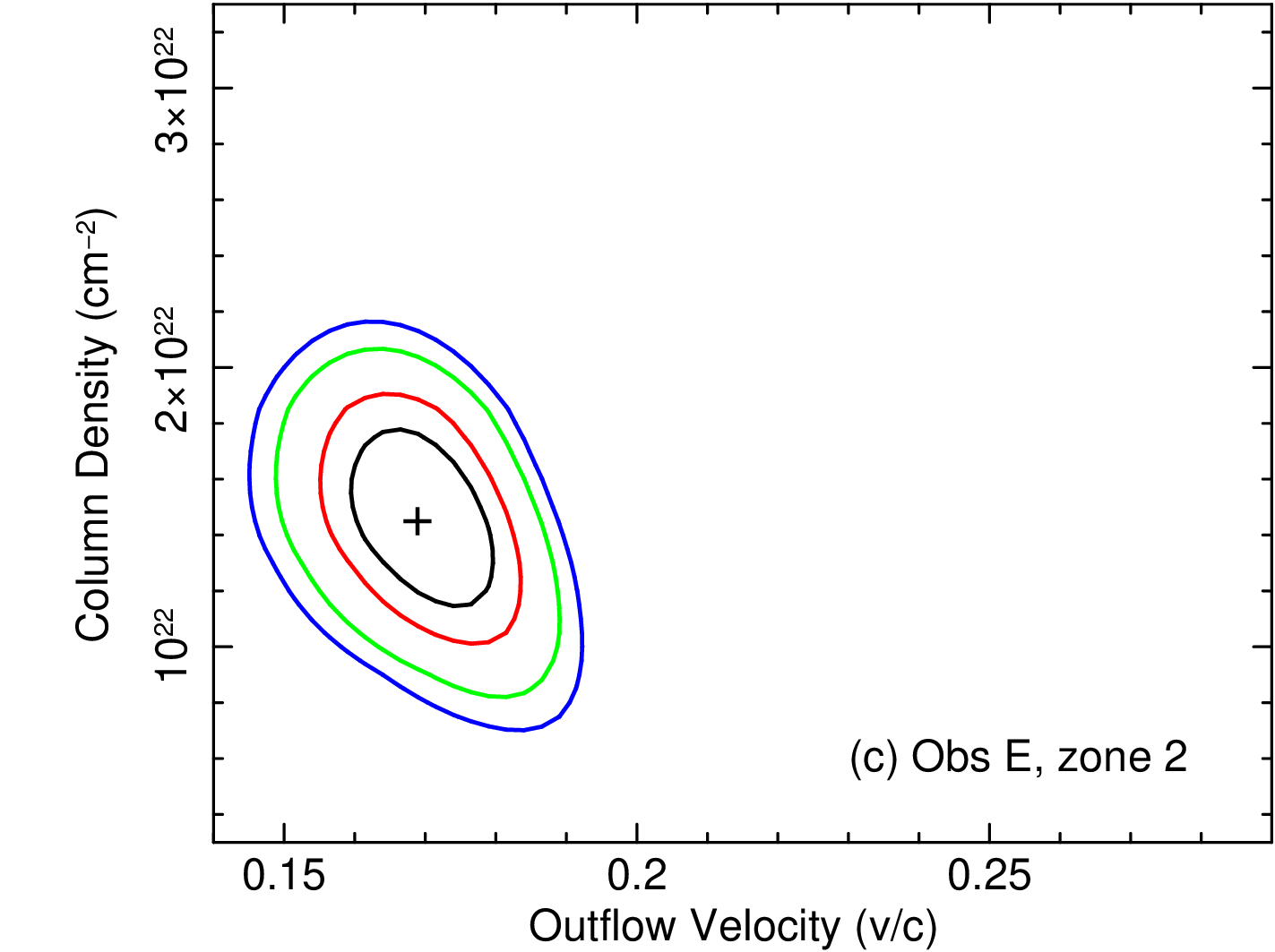}
\includegraphics[angle=-90,width=0.49\textwidth]{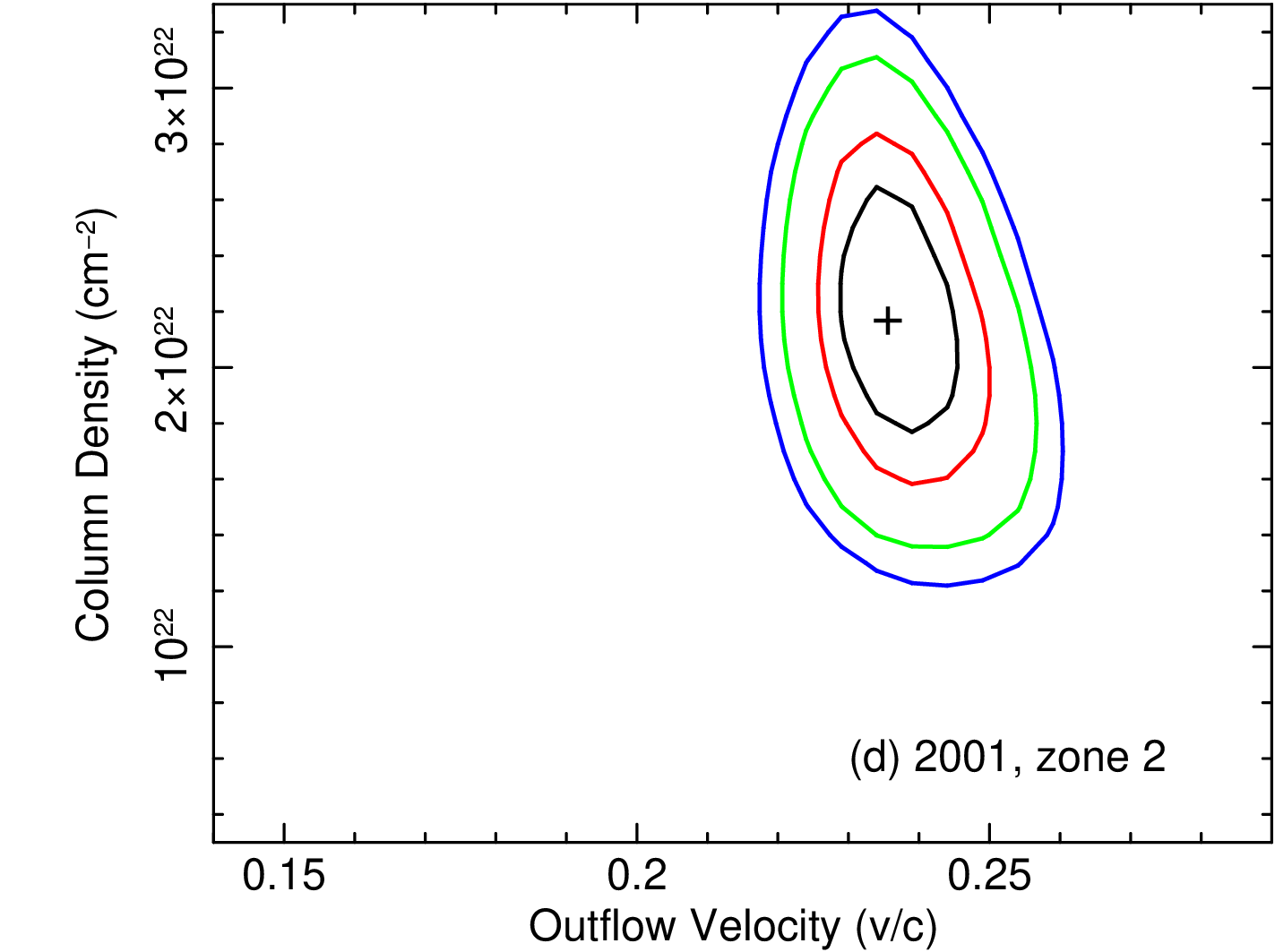}
\end{center}
\caption{Contour plots of column density against outflow velocity for the 
zone 2 absorber, for (a) OBS.\,B, (b) OBS.\,CD, (c) OBS.\,E, and (d) 2001. The 
inner (black) contour to the outermost (blue) contour represent the 68\%, 90\%, 
99\% and 99.9\% confidence contours for 2 parameters of interest, while the 
black crosses represents the best fit values. For OBS.\,B and OBS.\,CD, the 
confidence contours overlap, indicating that the soft 
X-ray absorber has not strongly varied on a 2 week timescale between these consecutive 
observations. However, on longer (months to years) timescales, the absorber 
clearly varies, e.g. between OBS.\,E (February 2014) and OBS.\,CD (September 
2013) the column has increased by at least a factor of two, while the velocity 
has decreased from  0.27\,$c$ to 0.17\,$c$. Across all of the observations, 
the absorber varies in both $N_{\rm H}$ and velocity, while the ionization 
parameter remains constant within errors. Note the identical $y$ and $x$ axis 
scales on all four plots.}
\label{contour}
\end{figure}

\clearpage

\begin{figure}
\begin{center}
\rotatebox{-90}{\includegraphics[width=12cm]{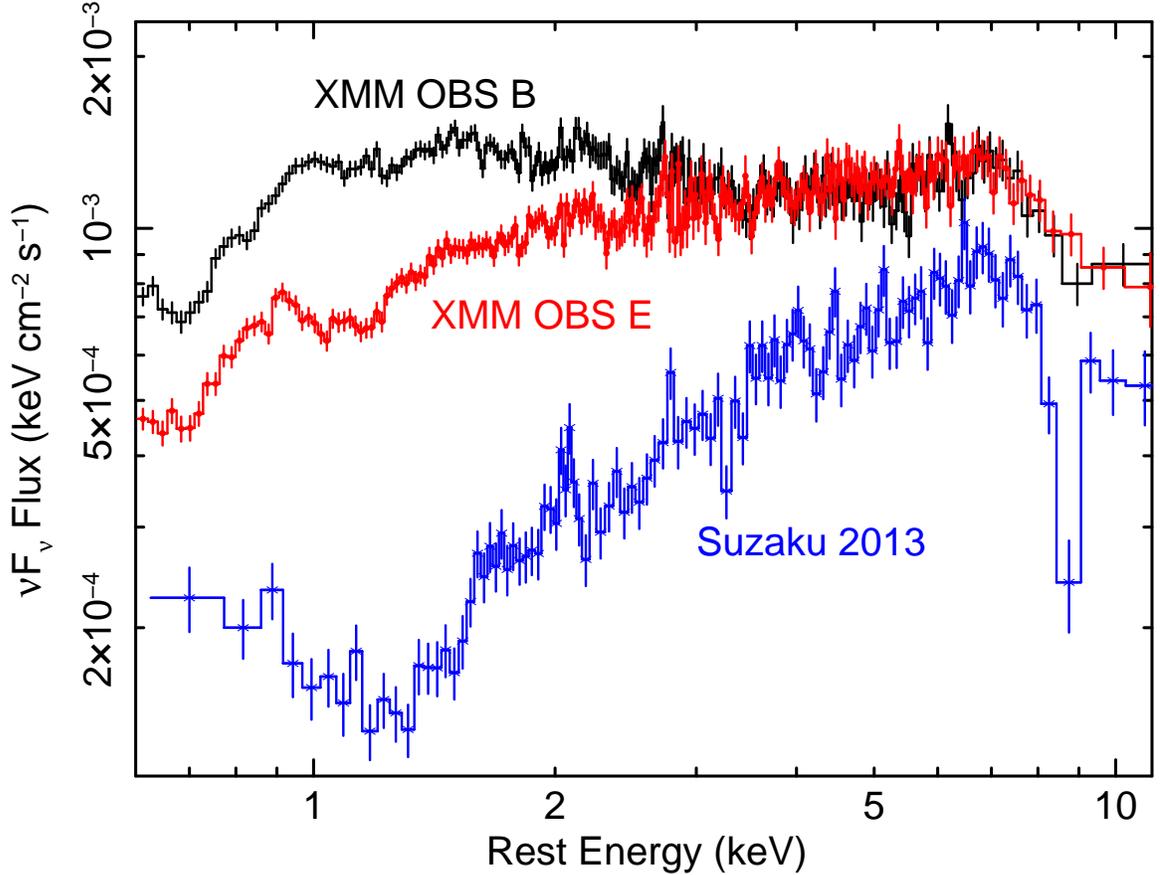}}
\end{center}
\caption{A comparison between the \xmm\ EPIC MOS spectra of OBS.\,B 
(September 2013, black crosses) and OBS.\,E (February 2014, red circles) 
versus the low flux \suzaku\ XIS\,0+3 spectrum of PDS\,456 obtained in March 2013 
(blue stars). The latter spectrum corresponds to the third sequence during the 
February--March 2013 \suzaku\ campaign on PDS\,456 (see Gofford et al. 2014, 
Matzeu et al. 2016). The $\nu F_{\nu}$ spectra have been fluxed against a 
$\Gamma=2$ power law. During the \suzaku\ 2013 observations the quasar was 
at an extreme low flux, with the spectrum heavily absorbed. During the OBS.\,B 
spectrum the quasar is relatively unobscured and the soft X-ray flux at 1\,keV 
has increased by a factor of $\times 10$ over a period of six months. Then 
during OBS.\,E, several months later, the hard X-ray flux has remained unchanged 
compared to OBS.\,B, while the soft X-ray flux below 2\,keV declined again by a 
factor of $\sim 2$, consistent with the increased level of absorption revealed 
by the RGS spectra.}
\label{compare}
\end{figure}

\begin{deluxetable}{lccccccc}
\tabletypesize{\small}
\tablecaption{Summary of PDS\,456 Observations}
\tablewidth{0pt}
\tablehead{
\colhead{Observation} & \colhead{ObsID} & \colhead{Instrument} & \colhead{Start Date/Time$^{\rm a}$} 
& \colhead{Exposure (ks)$^{\rm b}$} & \colhead{Net Rate (s$^{-1}$)$^{\rm c}$} & \colhead{Flux$^{\rm d}$}}

\startdata
%\hline\hline
\multicolumn{7}{c}{Current \xmm\ Observations} \\
\hline
2013 OBS.\,A & 0721010201 & RGS\,1+2 & 2013-08-27 04:23:26 & -- & --  & -- \\
& & MOS\,1+2 & & 105.8 & $1.176\pm0.003$ & 3.80\\
2013 OBS.\,B & 0721010301 & RGS\,1+2 & 2013-09-06 03:06:51 & 109.3 & $0.124\pm0.001$ & 2.52 \\
 		    &                      & MOS\,1+2 &  				   & 105.6 & $0.786\pm0.003$ & 2.58 \\
2013 OBS.\,C & 0721010401 & RGS\,1+2 & 2013-09-15 18:30:00 & 115.5 & $0.110\pm0.001$ & 2.28 \\
		  &			   & MOS\,1+2 & 	                            & 111.2 & $0.733\pm0.002$ & 2.42 \\
2013 OBS.\,D & 0721010501 & RGS\,1+2 & 2013-09-20 02:29:39 & 108.8 & $0.107\pm0.001$  & 2.25\\
			& 	           & MOS\,1+2 & 					& 107.7 & $0.719\pm0.002$  & 2.38\\
2014 OBS.\,E & 0721010601 & RGS\,1+2 & 2014-02-26 07:45:26 & 131.0 & $0.066\pm0.002$ & 1.58 \\
		    & 	  		  & MOS\,1+2 & 					 	& 114.9 & $0.538\pm0.002$ & 1.67 \\
Mean B--E & -- & RGS\,1+2 & -- & 464.6 & $0.102\pm0.001$ & 2.14 \\
\hline
%\hline\hline
\multicolumn{7}{c}{Archival \xmm\ Observations} \\
\hline
2001 & 0041160101 & RGS\,1+2 & 2001-02-26 09:44:02 & 43.4 & $0.209\pm0.003$ & 4.01 \\
	 & 	 		& MOS\,1+2 & 					& 43.7 & $1.315\pm0.005$ & 4.11 \\
2007 & 0501580101 & RGS\,1+2 & 2007-09-12 01:07:18 & 90.5 & $0.199\pm0.002$ & 4.25 \\
	 &		 	 & MOS\,1+2 & 					&87.6& $1.462\pm0.004$ & 4.42 \\
2007 & 0501580102 & RGS\,1+2 & 2007-09-14 01:46:07 & 89.1 & $0.126\pm0.002$  & 2.67 \\
	 & 	 		& MOS\,1+2 & 					& 86.1 & $0.950\pm0.003$  & 2.86
\\
\enddata

\tablenotetext{a}{Observation Start/End times are in UT.} 
\tablenotetext{b}{Net exposure time, after screening and deadtime correction, in ks.}
\tablenotetext{c}{Combined net count rate of RGS 1 and 2, or MOS 1 and 2 combined.} %or MEG 1st order.}
\tablenotetext{d}{Observed 0.5--2.0\,keV band fluxes, in units of $10^{-12}$\,erg\,cm$^{-2}$\,s$^{-1}$.}

\label{observations}
\end{deluxetable}

\clearpage

\begin{deluxetable}{lccccccc}
\tabletypesize{\small}
\tablecaption{ Best fit parameters of the strongest Gaussian lines fitted to RGS.}
\tablewidth{0pt}
\tablehead{
\colhead{ }  & \colhead{Rest Energy$^{\rm a}$}   &$\sigma$  & $\sigma_{\rm v}$  &  \colhead{Intensity}  &  \colhead{EW} &\colhead{$\Delta \chi^2/\Delta\nu$} & \colhead{$P_{F}$$^{b}$}\\
\colhead{ }  & \colhead{(eV)} & (eV) & \colhead{km s$^{-1}$}& \colhead{($10^{-5}$ ph cm$^{-2}$ s$^{-1}$)}&    \colhead{(eV)} &   
}
\startdata    
\hline
\hline
\multicolumn{7}{c}{OBS.\,CD: No Gaussian lines,  $\chi^2/\nu=170.2/100$, $N_{\rm P}=1.53\times10^{-5}$$^{c}$} \\
\hline
ABS1 &$846\errUD{6}{6}$ 	& $6.5^{+6.0}_{-2.8}$  &	$2,300^{+2,100}_{-1,000}$	&$-5.0\errUD{2.4}{2.4}$	& $-4.6\errUD{2.2}{2.2}$ &17.0/3 & $8.2\times10^{-3}$ \\
ABS2 &$1174\errUD{47}{58}$ 	& $109\errUD{49}{35}$  & $28,000^{+13,000}_{-9,000}$	&	$-13.0\errUD{4.0}{4.0}	$& $-41\errUD{13}{13}$ &	25.2/3 & $7.1\times10^{-4}$\\
%EMIS1 &$792\errUD{8}{8}$ 	& $6.5^{\mathrm{t}}$   &		&$6.7\errUD{3.2}{3.2}	$& $4.8\errUD{2.3}{2.3}$ &	11.2\\
&&&&&& \\
\hline \hline
\multicolumn{7}{c}{OBS.\,E:  No Gaussian lines,  $\chi^2/\nu=149.9/100$, $N_{\rm P}=9.2\times10^{-4}$$^c$} \\
\hline
ABS1 &$1016\errUD{19}{18}$ 	& $41\errUD{14}{12}$  &	$12,000^{+4,000}_{-3,500}$ &$-11.2\errUD{8.0}{5.4}$	& $-27.3\errUD{19.3}{13.2}$ & 17.3/3 & $1.4\times10^{-3}$\\
ABS2 &$1166\errUD{19}{20}$ 	& $41^{\rm t}$  &		&$-6.7\errUD{4.1}{2.8}	$& $-31.3\errUD{19.1}{13.0}$ &	27.0/2 & $9.0\times10^{-6}$\\
EMIS1 &$913\errUD{16}{15}$ 	& $18\errUD{13}{8}$  &	$6,000^{+4,000}_{-3,000}$	&$5.2\errUD{4.3}{3.2}	$& $9.8\errUD{8.1}{6.0}$ &	11.6/3 & 0.013\\
%EMIS2 &$1051\errUD{11}{10}$ 	& $18^{\mathrm{t}}$ &		&$4.8\errUD{2.7}{2.0}	$& $20.1\errUD{11.0}{8.5}$ &	4.3/2\\
&&&&& &\\
\hline \hline
\multicolumn{7}{c}{2001: No Gaussian lines,  $\chi^2/\nu=181.5/99$, $N_{\rm P}=8.4\times10^{-7}$$^c$} \\
\hline 
%ABS1 &$925\errUD{6}{6}$ 	& $9.5^{+7.3}_{-4.6}$  &	$3100^{+2400}_{-1500}$	&$-15.0\errUD{6.6}{-8.3}$	& $-6.8\errUD{3.0}{3.8}$ & 14.2/3\\
ABS1 &$1061\errUD{11}{11}$ 	& $42\errUD{13}{11}$  & $11,900^{+3,700}_{-3,100}$	&	$-40.6\errUD{12.4}{16.8}	$& $-37\errUD{11}{15}$ &	42.0/3 & $1.32\times10^{-5}$\\
\enddata
\tablenotetext{a}{Measured line energy in the PDS\,456 rest frame.}   \\
\tablenotetext{b}{Null hypothesis probability of any individual line, via F-test}
\tablenotetext{c}{Fit statistic and null hypothesis probability against baseline continuum model (without Gaussian lines).}
\tablenotetext{t}{Denotes a tied parameter.}
%$^{\mathrm{e}}$ velocity in km\,s $^{-1}$.\\
%$^f$ Denotes the parameter is fixed.}
\label{Gauss}
\end{deluxetable}

\clearpage
\begin{deluxetable}{lcccccc}
%\tabletypesize{\scriptsize}
\tablecaption{ Best fit  parameters of the strongest absorption lines detected in the MOS spectra.}
\tablewidth{0pt}

\tablehead{
\colhead{ }  & \colhead{Rest Energy$^{\rm a}$}   &  \colhead{Intensity}  &  \colhead{EW} & \colhead{$\sigma$} & \colhead{$\Delta \chi^2/\Delta\nu$} & \colhead{$P_{F}$$^{b}$}\\
 \colhead{ }  & \colhead{(eV)}  & \colhead{($10^{-5}$ ph cm$^{-2}$ s$^{-1}$)}&    \colhead{(eV)} &  \colhead{(eV)} 
 }
 \startdata    
\hline\hline
\multicolumn{7}{c}{OBS.\,B: No Gaussian lines, $\chi^2/\nu=172.3/101$, $N_{\rm P}=1.3\times10^{-5}$$^c$} \\
\hline
ABS1 & $ 839	\errUD{ 16}{4 }$ & $-8.2\errUD{ 4.2}{3.5 }$ & $-7.6	\errUD{4.0 }{3.2 }$ & $10^{\rm f}$ & 18/2 & $1.2\times10^{-3}$\\
ABS2 &  $	1253\errUD{18 }{18 }$ 	& $-2.1	\errUD{ 0.7}{ 0.7}$  	&$-7.7 	\errUD{2.5 }{2.6 }$   & $100^{\rm f}$ & 34.1/2 & $5.5\times10^{-6}$ \\

&&& &\\
\hline
\hline
\multicolumn{7}{c}{OBS.\,CD: No Gaussian lines, $\chi^2/\nu=198.1/102$, $N_{\rm P}=3.8\times10^{-8}$$^c$} \\
\hline
ABS1 &$	845\errUD{ 16}{15 }$ &$	-4.7\errUD{ 2.5}{2.3 }$ &$-4.9	\errUD{2.6}{2.4 }$ & $10^{\rm f}$ & 13.6/2 & $3.9\times10^{-3}$\\
 ABS2 &$1249	\errUD{ 21}{ 21}$ &$	-6.8\errUD{1.3 }{1.3 }$ &$-24.2\errUD{4.6 }{4.6 }$ & $110^{\rm f}$ & 71.0/2 & $4.6\times10^{-11}$\\
&&& &\\
\hline
\hline
\multicolumn{7}{c}{OBS.\,E: No Gaussian lines, $\chi^2/\nu=208.3/101$, $N_{\rm P}=1.9\times10^{-9}$$^c$} \\
\hline
ABS1 &$	998\errUD{ 17}{16 }$&$-6.2	\errUD{2.1 }{ 1.8}$&$-16.8\errUD{ 5.7}{5.2 }$ & $45^{+30}_{-15}$ & 38.7/3 & $5.8\times10^{-6}$\\
ABS2 &$	1172\errUD{13 }{ 14}$&$-4.0	\errUD{0.9 }{0.9 }$&$-19.2	\errUD{4.4 }{4.2 }$ & $45^{\rm t}$ & 33.4/2 & $7.4\times10^{-6}$\\
EMIS1 & $933\pm23$ & $3.2^{+1.7}_{-1.6}$ & $7.1^{+3.1}_{-3.0}$ & $10^{\rm f}$ & 44.7/2 & $2.4\times10^{-7}$\\ 
&&& &\\
\hline
\hline
\multicolumn{7}{c}{2001: No Gaussian lines, $\chi^2/\nu=260.3/100$, $N_{\rm P}=3.16\times10^{-16}$$^c$} \\
\hline
ABS1 &$1083	\errUD{8 }{8 }$&$-21.1	\errUD{3.7 }{3.2 }$&$-32.2	\errUD{5.7 }{4.9 }$ & $47^{+11}_{-10}$ & 79.1/3 & $1.8\times10^{-11}$\\
ABS2 &$	1254\errUD{ 12}{13 }$&$-8.6	\errUD{ 1.7}{1.6 }$&$-22.9	\errUD{ 4.6}{ 4.2}$ & $47^{\rm t}$ & 74.7/2 & $1.1\times10^{-11}$\\
&&& &\\
%\hline
%\hline
%\multicolumn{5}{c}{2007} \\
%\hline
%ABS1 &$1104	\errUD{20 }{19 }$&$	-3.14\errUD{1.0 }{0.93 }$&$-6.5	\errUD{2.1 }{1.9 }$ &34.6\\
%&&& &\\
%
 \enddata
\tablenotetext{a}{ Measured line energy in the PDS\,456 rest frame.} \\
\tablenotetext{b}{Null hypothesis probability of any individual line, via F-test}
\tablenotetext{c}{Fit statistic and null hypothesis probability against baseline continuum model (without Gaussian lines).}
\tablenotetext{f}{Denotes a fixed parameter. Where the line width is unresolved in the MOS, 
it is fixed at a width of $\sigma=10$\,eV. For the broad (ABS\,1) profile in OBS.\,CD, the 
width of $\sigma=110$\,eV is fixed to the line width obtained from the RGS spectrum.} 
\tablenotetext{t}{Denotes a width which is tied between the two pairs of absorption lines.} 
\label{MOSgauss}
\end{deluxetable}

\clearpage

\begin{deluxetable}{ccccccc}
\tablecaption{Best fit parameters for the ionized absorbers.}
\tablewidth{0pt}
\tablehead{
\colhead{}  & &  \colhead{B} & \colhead{CD} &\colhead{E} & \colhead{2001} &  \colhead{ 2007}  \\
 }
 \startdata    

Zone 1  & $\nhsym ^{\rm a}$ &$-$&$-$&$0.61\errUD{0.25}{0.20}$&$1.9\errUD{0.5 }{0.5 }$&$-$\\
 & $\log \xi ^{\rm b}$ & $-$&$-$	&$3.65	\errUD{0.13 }{0.13 }$&$4.09\errUD{0.07 }{ 0.10}$ &$-$\\
& $v_{\mathrm{out}}/c$  &$-$	&	$-$&$-0.064 	\errUD{0.012 }{0.012 }$	 	&$-0.082 \errUD{0.01 }{0.01 }$&$-$\\
 & $\Delta \chi^2$ &	$-$&$-$	&24.9&31.8 &$-$\\

\hline
\hline
Zone 2  & $\nhsym ^{\rm a}$ &$0.52\errUD{0.29 }{ 0.20}$	& $0.70	\errUD{ 0.22}{0.21}$&$1.50\errUD{0.40}{0.45}$&$2.2	\errUD{ 0.7}{0.6 }$ &$<0.23$  \\
 & $\log \xi^{\rm b}$  &	$4.04\errUD{0.13 }{ 0.29}$&	$	4.18\errUD{ 0.05}{ 0.08}$&$4.18	\errUD{0.08 }{0.11 }$&$	4.19\errUD{0.07 }{0.08 }$ &$	4.10^f$\\
& $v_{\mathrm{out}}/c$  &$-0.254	\errUD{ 0.011}{0.013 }$	&$	-0.267\errUD{0.006}{ 0.007}$	&$	-0.170\errUD{0.012 }{0.011 }$& $-0.237\errUD{0.013 }{0.013 }$ & $-0.177\errUD{ 0.015}{0.011 }$\\
 & $\Delta \chi^2$ &	50.4&	80.5&83.0&20.2 &6.7\\
\hline
\hline
Zone 3  & $\nhsym ^{\rm a}$ & $<31$	& $23 \errUD{ 8}{8}$ & $<37$ & $<46$  &$<22$ \\
 & $\log \xi^{\rm b}$  & $5.5^{\rm f}$	& $5.5^{\rm f}$& $5.5^{\rm f}$ & $5.5^{\rm f}$ & $5.5^{\rm f}$\\
& $v_{\mathrm{out}}/c$  & $-0.3^{\rm f}$ & $-0.30\errUD{0.01}{ 0.01}$	& $-0.3^{\rm f}$ & $-0.3^{\rm f}$ & $-0.3^{\rm f}$ \\
 & $\Delta \chi^2$ & -- & 20.0	& -- & -- & 10.9\\

  \enddata
\tablenotetext{a}{Column density in units of $10^{22}$ \nh . Note that for OBS.\,A, 
the upper-limit on the column density (EPIC only) is $<3.5\times10^{21}$\,cm$^{-2}$ for 
an ionization parameter of $\log \xi=4.1$.} \\
\tablenotetext{b}{Ionization parameter in units of \logxi.}
\tablenotetext{f}{Denotes a fixed parameter.}

\label{xstar_sequences}
\end{deluxetable}

\clearpage

\begin{deluxetable}{lcc}
\tabletypesize{\small}
\tablecaption{Photoionized Emission Parameters for Observation E.}
\tablewidth{0pt}
\tablehead{
\colhead{Parameter} & \colhead{Zone 1 (low ionization)} & \colhead{Zone 2 (high ionization)}}

\startdata
\hline
$N_{\rm H}$$^{\rm a}$ & 0.6$^{\rm f}$ & 1.5$^{\rm f}$ \\
$\log\xi$$^{\rm b}$ & $2.8\pm0.3$ & $4.6^{+0.6}_{-0.4}$ \\
$v_{\rm out}/c$ & $-0.040\pm0.018$ & $-0.08\pm0.02$ \\
$\kappa_{\rm xstar}$$^{\rm c}$ & $6.0\pm3.0 \times10^{-4}$ & $<1.2\times10^{-3}$\\
$f=\Omega/4\pi$$^{\rm d}$ & $0.44\pm0.22$ & $<0.88$\\
\enddata
\tablenotetext{a}{Column density in units of $10^{22}$ \nh. Note the column is tied 
to the respective absorption zone.} 
\tablenotetext{b}{Ionization parameter in units of \logxi.}
\tablenotetext{c}{The normalization of the {\sc xstar} component, which is defined 
as $\kappa=fL_{38}/D_{\rm kpc}^{2}$, where $L_{38}$ is the ionizing luminosity in 
units of $10^{38}$\,erg\,s$^{-1}$, $D_{\rm kpc}$ is the source distance in kpc and 
$f$ is the covering fraction of the photoionized gas. See Section 6.3 for details.}
\tablenotetext{d}{Covering fraction, where $f=\Omega/4\pi$.}
\tablenotetext{f}{Denotes a fixed parameter.}
\label{photoionized}
\end{deluxetable}

\end{document}